\documentclass[a4paper,manuscript]{aastex}

\usepackage{lineno}

\usepackage{amssymb,amsmath}
\usepackage{gensymb}
\usepackage{pifont}
\usepackage{graphicx}
\usepackage{multirow}
\usepackage[dvipsnames]{xcolor}

\usepackage{pdfpages}
\usepackage[pdfpagelabels=true, pdftitle={Statistics of 700 individually studied W UMa stars}, pdfauthor={Latkovi{\' c} et al.}, ]{hyperref}
\usepackage{bookmark}

\def \objCount {688 }
\def \oldCount {237 }
\def \bibCount {462 }
\def \newCount {478 }

\def \qSPCount {159}
\def \qPHCount {529 }
\def \PECount {371 }
\def \TECount {317 }
\def \qSPfracTE {26}
\def \qSPfracPE {34}
\def \ATCount {176 }
\def \WTCount {225 }
\def \AWCount {401 }
\def \cATCount {350 }
\def \cWTCount {338 }
\def \dAWCount {62 }

\def \AmedP {0.40}
\def \WmedP {0.31}
\def \cAmedP {0.39}
\def \cWmedP {0.32}
\def \AmedF {0.35}
\def \WmedF {0.16}
\def \cAmedF {0.25}
\def \cWmedF {0.18}
\def \AmedQ {0.30}
\def \WmedQ {0.43}

\def \BCount {31 }
\def \HCount {118 }
\def \HCountRel {16 }

\def \SPCount {343 }
\def \USCount {345 }

\def \HL3Count {100 }
\def \NL3Count {588 }
\def \L3Count {688 }
\def \dPCount {271 }

\def \PincCount {120 }
\def \PdecCount {98 }
\def \PcCount {135 }
\def \dPcCount {98 }

\def \Palambda {0.82 }
\def \Pblambda {0.18 }
\def \Pamu {0.33 }
\def \Pbmu {0.62 }
\def \Pasigma {0.08 }
\def \Pbsigma {0.19 }
\def \Pmin {0.18 }

\def \Ploq {0.28 }
\def \Pmed {0.35 }
\def \Pupq {0.43 }

\def \objPmin {CSS J214633.8+120016 }

\def \objPus {27 }

\def \dPmu {0.0055 }
\def \dPsigma {0.0278 }

\def \dPloq {-0.01 }
\def \dPmed {0.00 }
\def \dPupq {0.02 }

\def \objdPmax {V530 And }
\def \objdPmaxVal {-1.52928 }
\def \PCmu {1.27 }
\def \PCsigma {0.39 }
\def \PCmin {1.77 }

\def \PCloq {10.54 }
\def \PCmed {19.70 }
\def \PCupq {38.18 }

\def \objPCmin {V345 Gem }

\def \Qalambda {0.81 }
\def \Qblambda {0.19 }
\def \Qamu {0.34 }
\def \Qbmu {0.83 }
\def \Qasigma {0.15 }
\def \Qbsigma {0.10 }
\def \Qmin {0.04 }

\def \Qloq {0.25 }
\def \Qmed {0.39 }
\def \Qupq {0.55}

\def \objQmin {V1187 Her }

\def \fZeroCount {5 }
\def \Falambda {0.58 }
\def \Fblambda {0.42 }
\def \Famu {0.14 }
\def \Fbmu {0.50 }
\def \Fasigma {0.08 }
\def \Fbsigma {0.23 }

\def \Floq {0.11 }
\def \Fmed {0.20}
\def \Fupq {0.43}

\def \Fmax {0.9997 }
\def \objFmax {V345 Gem }
\def \FZObjs {1SWASP J011732.10+525204.9, 44 Boo, KIC 1572353, MX Peg, NSVS 4316778}
\def \Ialambda {0.58 }
\def \Iblambda {0.42 }
\def \Iamu {81.38 }
\def \Ibmu {67.21 }
\def \Iasigma {4.79 }
\def \Ibsigma {11.79 }

\def \Iloq {69.8 }
\def \Imed {78.2 }
\def \Iupq {82.9 }

\def \Tmu {5758 }
\def \Tsigma {928 }
\def \Tmin {3600 }

\def \Tloq {5072 }

\def \Tupq {6340 }

\def \Tmax {9600 }
\def \objTmin {CSS J214633.8+120016 }
\def \objTmax {AU Pup }

\def \TDamu {-39 }

\def \TDasigma {251 }

\def \TDloq {-205 }
\def \TDmed {0 }
\def \TDupq {176 }

\def \RCount {407 }
\def \rimu {0.45 }
\def \rzmu {0.29 }
\def \risigma {0.05 }
\def \rzsigma {0.05 }
\def \mCount {437 }
\def \mimu {0.07 }
\def \mzmu {-0.42 }
\def \misigma {0.16 }
\def \mzsigma {0.25 }
\def \Mmax {4.92 }
\def \Mmin {0.60 }
\def \objMmax {UCAC2 31686238}
\def \objMmin {CRTS J130945.0+371627}
\def \lCount {408 }
\def \limu {0.21 }
\def \lzmu {-0.24 }
\def \lisigma {0.54 }
\def \lzsigma {0.43 }
\def \Aalambda {0.81 }
\def \Ablambda {0.19 }
\def \Aamu {2.35 }
\def \Abmu {4.07 }
\def \Aasigma {0.52 }
\def \Absigma {0.84 }
\def \ACount {441 }
\def \Amin {0.95 }

\def \Aloq {2.07 }

\def \Aupq {3.00 }

\def \Amax {6.16 }
\def \objAmin {EE Cet }
\def \objAmax {KIC 11097678 }
\def \Gmu {4.52 }
\def \Gsigma {1.68 }
\def \GCount {313 }

\def \Gloq {3.42 }

\def \Gupq {6.51 }

\def \objGmin {V816 Cep }
\def \objGminG {0.51 }
\def \TEmedianQ {0.3 }
\def \PEmedianQ {0.5 }
\def \TEvarQ {0.03}
\def \PEvarQ {0.06}
\def \lamostCount {5256 }
\def \crtsCount {9380 }

\def \qTok {475 }

\def \qMok {299 }

\def \qRok {275 }

\def \qLok {216 }

\def \PTok {591 }

\def \PMok {235 }

\def \PRok {222 }

\def \PLok {210 }

\def \MRok {277 }
\def \MLok {278 }

\shorttitle{Statistics of W UMa stars}
\shortauthors{Latkovi{\' c} et al.}

\begin{document}

\title{Statistics of 700 individually studied W UMa stars}

\author{Olivera Latkovi{\' c}}
\affil{Astronomical Observatory, Volgina 7, 11060 Belgrade, Serbia}
\email{olivia@aob.rs}

\author{Atila {\v C}eki}
\affil{Astronomical Observatory, Volgina 7, 11060 Belgrade, Serbia}

\and

\author{Sanja Lazarevi{\' c}}
\affil{Astronomical Observatory, Volgina 7, 11060 Belgrade, Serbia}

\begin{abstract}
We present a statistical study of the largest bibliographic compilation of stellar and orbital parameters of W UMa stars derived by light curve synthesis with Roche models. The compilation includes nearly 700 individually investigated objects from over 450 distinct publications. Almost 70\% of this sample is comprised of stars observed in the last decade that have not been considered in previous statistical studies. We estimate the ages of the cataloged stars, model the distributions of their periods, mass ratios, temperatures and other quantities, and compare them with the data from CRTS, LAMOST and Gaia archives. As only a small fraction of the sample has radial velocity curves, we examine the reliability of the photometric mass ratios in totally and partially eclipsing systems and find that totally eclipsing W UMa stars with photometric mass ratios have the same parameter distributions as those with spectroscopic mass ratios. Most of the stars with reliable parameters have mass ratios below 0.5 and orbital periods shorter than 0.5 days. Stars with longer periods and temperatures above 7000 K stand out as outliers and shouldn't be labeled as W UMa binaries. 

The collected data is available as an online database at \textit{https://wumacat.aob.rs}.
\end{abstract}

\keywords{binaries: eclipsing -- binaries: close -- stars: fundamental parameters}

%
%

\section{Introduction}
\label{secIntro}

W UMa stars are late-type contact binaries with short periods (of about 0.35 days) whose components typically have very different masses (with mass ratios around 0.3) but nearly identical temperatures ($\Delta T \le 200 K$). This paradoxical property is caused by the exchange of mass and energy through the common envelope \citep[see e.g.][]{webbink, YE2005}. The progenitors of W UMa stars are low-mass close binaries that evolve into contact through the shrinking of the orbit caused by angular momentum loss from magnetic breaking \citep{stepien2011}. The presence of a gravitationally bound third body may also play a role, as most W UMa binaries are suspected, and many confirmed to be members of hierarchical multiple systems \citep{PR2006, li2018}.

Despite the abundance of observational and theoretical studies in the literature, the understanding of W UMa stars is far from complete. Seminal works of \citet{rucinski2015, rucinski2020} where large telescopes were used to obtain high-resolution time-resolved spectroscopy of two bright W UMa stars revealed large-scale surface flows and other phenomena unaccounted for by the widely accepted model of \citet{lucy68a, lucy68b}. Based on the numerical simulations of accretion flow in semi-detached binaries done by \citet{oka2002}, \citet{stepien2009} proposed a model of W UMa binaries where the energy exchange between the components is achieved through large-scale circulation in the equatorial plane, essentially predicting the observational results of \citet{rucinski2015, rucinski2020}, mentioned above. The same study suggests that W UMa binaries may have gone through the mass ratio reversal and are in a similar evolutionary stage as low-mass Algols, a scenario which is at odds with the generally accepted view. 

The current work aims to contribute to the understanding of these fascinating stars by summarizing hundreds of recent observational studies.

Thanks to the small orbital separations, W UMa stars often display eclipses; at the same time, short periods make them attractive for observations with smaller ground-based instruments. Hundreds of these stars have been studied using the light curve synthesis approach that has been made widely accessible by the advent of the Wilson-Devinney (henceforward WD) modeling code \citep{wd1971, wd2008, wd2020ascl}, which remains the most popular software for this kind of work to date (see Section \ref{secOurCat}). This and other modeling tools based on the Roche formalism can calculate the orbital inclination, radii of stars relative to the orbital separation, and the ratio of their temperatures from the light curve. The estimates of the mass ratio, orbital separation (ideally from radial velocity curves), and the temperature of the brighter component then allow the determination of the so-called absolute stellar parameters: the masses, radii, temperatures and luminosities of the components in solar units. The details of this process are beyond the scope of this study, but the readers can find them in, say, \citet{kallrath-milone}.

While most of the above is true for all eclipsing binaries, W UMa stars have additional peculiarities that make them especially interesting. Having a common envelope, the components share the same Roche surface, and as a consequence, the ratio of their radii depends on the the mass ratio as $R_2/R_1\approx (M_2/M_1)^{0.46}$ \citep{kuiper41}. Since the radius ratio is easily obtained from the light curve model (when there's a total eclipse), so is, in principle, the mass ratio---even in the absence of radial velocity measurements.

Another consequence of the common envelope is that a period-luminosity relation can be derived (and has been observed) for W UMa stars, making them viable distance tracers \citep{rucinski2004}. \citet{Chen2016} studied a large sample of contact binaries in open clusters and showed that these stars are competitive distance indicators with uncertainties better than 5\% for 90\% of the sample. Based on absolute magnitudes of several hundred W UMa stars from the Tycho-Gaia Astrometric Solution parallax data, \citet{Mateo2017} found a steep linear relationship between the absolute magnitude and the orbital period that can serve as a calibration in the period range from 0.22 to 0.88 days over five magnitudes in $M_v$.

With the present and future deluge of photometric time series data from ground-based surveys and space telescopes, these properties of W UMa stars are growingly valuable.

Several groups compiled the published parameters of W UMa stars and performed various analyses prior to this work. \citet{MV1996} collected from the literature the light curve solutions\footnote{A ``light curve solution'' is a collection of binary system model parameters that can be used to calculate a synthetic light curve which fits the observations.} of 78 contact binaries, derived their absolute parameters and conducted a statistical analysis. Almost a decade later, \citet{PR2003} and \citet{CK2004} published similar compilations that increased the joint sample of cataloged W UMa stars with complete solutions to about 200. \citet{YE2005} studied a portion of this sample (contributing several new systems) and pointed out the inconsistencies of light curve solutions found in the literature, such as cited spectral types often being inconsistent with the cited temperatures and cited radii often being smaller than the volume radius of the Roche lobes. \citet{GS2008} compiled a sample of 112 well-studied W UMa binaries with absolute parameters based on high-quality light and radial velocity curves and derived empirical relations that enable estimating all the key absolute parameters of a contact binary from its period. \citet{DS2011} modeled the All Sky Automated Survey \citep{ASAS} light curves of 62 eclipsing binaries with mass ratios previously established in spectroscopic studies, and published their stellar and orbital parameters, contributing a dozen new systems to the joint sample of W UMa stars.

\begin{figure}
\caption{Left: W UMa stars presented in different catalogs to date. References are available in Table \ref{tabOldCats}. ``Total'' is the number of complete light curve solutions (as defined in Section \ref{secOurCat}) for W UMa stars; detached and semidetached binaries, as well as early-type contact binaries are omitted from the counts. The ``New" objects are those that haven't been considered in previous compilations. Right: The number of light curve solutions in our catalog by solver.}
\label{figCatSol}
\includegraphics[width=.9\textwidth]{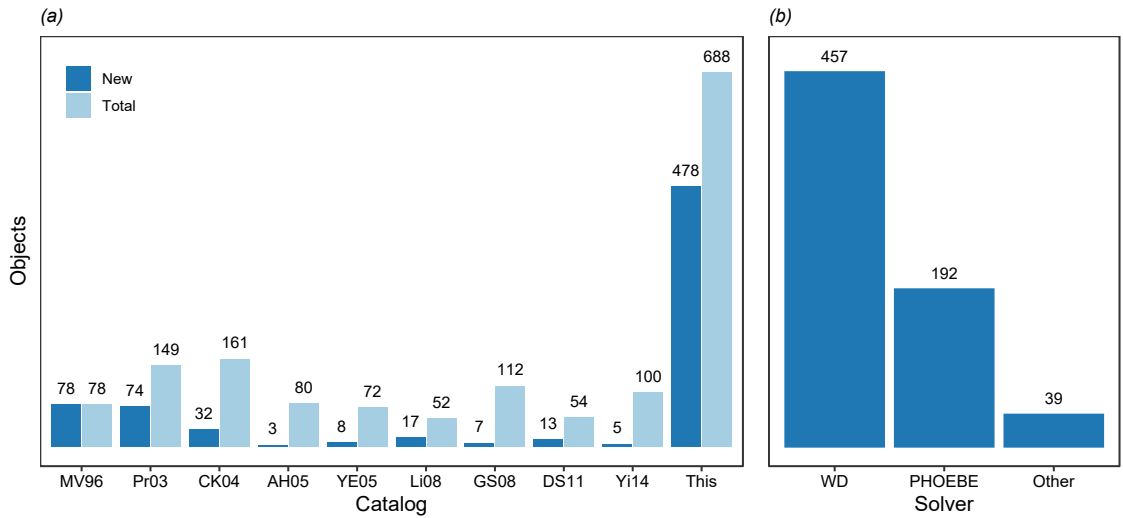}
\end{figure}

At the time of this writing, a decade has passed since the last of these publications. The catalogs mentioned above jointly contain information for \oldCount distinct objects with light curve solutions (we will refer to these stars as the ``old sample"). However, the studies of W UMa stars have not ceased; to the contrary, the number and quality of publications presenting light curves and models both for well-studied and neglected objects has been steadily increasing. After reviewing \bibCount relevant papers, we compiled a catalog of \objCount W UMa stars, of which \newCount have not been considered in any of the previous statistical studies. Fig. \ref{figCatSol}a shows the relative contribution of this ``new sample'' with regards to the compilations comprising the old sample. Our catalog of individually studied W UMa stars is the most exhaustive to date. 

\begin{table}
    \begin{center}
    \caption{Previous catalogs}
    \label{tabOldCats}
    \vspace{10pt}
    \begin{tabular}{lll}
    \tableline\tableline
    Fig. \ref{figCatSol} Label & Reference \\ 
    \tableline
    MV96 & \citet{MV1996} \\
    Pr03 & \citet{PR2003} \\
    CK04 & \citet{CK2004} \\
    AH05 & \citet{AH2005} \\
    YE05 & \citet{YE2005} \\
    Li08 & \citet{LI2008} \\
    GS08 & \citet{GS2008} \\
    DS11 & \citet{DS2011} \\
    Yi14 & \citet{YI2014} \\
    \tableline
    \end{tabular}
    \end{center}
\end{table}

It is also the most complete in terms of cataloged information. We use the available data to estimate the ages of the objects (Section \ref{secAge}) and calculate missing quantities whenever possible (Section \ref{secMQ}).

The entire catalog, with additional data from the Catalina Real-Time Transient Survey (CRTS) Variable Sources Catalogue \citep{crts}, the LAMOST catalog of contact binaries \citep{lamost} and the Gaia archive \citep{gaia1,gaia2}, is available as an online database at \textit{https://wumacat.aob.rs}. Among other functions, the website provides a handy tool for listing catalog stars similar to user-supplied parameters.

In what follows, we describe the data collection and the structure of the catalog. Sections \ref{secStats} and \ref{secRel} detail the statistical analyses we performed on the collected data.

%
%

\section{Catalog Description}
\label{secOurCat}

The contents of the catalog is shown in Table \ref{tabColumns}, which lists all the cataloged quantities with brief explanations. To permit a light curve solution into the catalog, we followed these rules:
\begin{itemize}
    \item The solution must indicate an exact contact or overcontact configuration. We omit near-contact solutions.
    \item The solution must be based on a Roche model. This disqualifies various ``phenomenological" methods, such as fitting the light curve with trigonometric or other analytical functions, machine learning methods and so forth. For a list of Roche models featuring in the catalog, see Table \ref{tabSolvers}. Most of the cataloged solutions were done with various versions of the WD code \citep{wd1971, wd2008, wd2020ascl}; the next most numerous group are solutions derived with PHOEBE \citep{phoebe2005, phoebe2011ascl}. All the other solvers are only represented by a small number of cases (see Fig. \ref{figCatSol}b).
    \item We also do not include the results of automated modeling of large collections of objects, such as those reported by \citet{sun2020} and \citet{li2020}, because our focus is on individual studies.
    \item The solution must at the very least contain the orbital period, the mass ratio, the temperatures of the components, the orbital inclination and either the fillout, or the value of the dimensionless Roche potential of the common envelope as a measure of relative stellar sizes (see also Section \ref{secMQ}). A full list of model parameters with definitions is given in Table \ref{tabColumns}.
    \item The synthetic light curve must fit the observations well. We omit the solutions where, for example, the light curves are asymmetric but the model doesn't account for this by including spots, or where the observations clearly show a total eclipse but the synthetic light curves have partial eclipses.
    \item We omit the systems where one or both components are of an early spectral type (B or earlier). The distinction is made according to the effective temperature, with the cut at $T=10,000 K$ \citep[see e.g.][]{allen}.
\end{itemize}

Most of the stars in the old sample and a fair portion of the new sample have more than one solution in the literature. We usually picked the newest solution, unless an older one was more complete, had a superior fit to the observations, or utilized higher quality data. Solutions based on combined radial velocity and light curve analysis (or those where the light curve synthesis relies on the results of previous spectroscopic studies) were preferred to those based on light curve modeling alone.

Many W UMa stars have dedicated period change studies in separate publications. When a period change study is done in the same paper with the light curve synthesis, we use those results; otherwise we search the literature for an independent period change study and catalog that as well.

We consider the mass ratio to be spectroscopic if the initial or final value originates from a radial velocity study; and photometric if no radial velocities are available or were not considered. The fillout is defined as $f=(\Omega_{in}-\Omega)/(\Omega_{in}-\Omega_{out})$ and expressed as a fraction of unity ($0 \le f \le 1$); $\Omega_{in}$ and $\Omega_{out}$ are the values of the dimensionless Roche potential at the Lagrange points $L_1$ and $L_2$, respectively.

\begin{table}
\begin{center}
\caption{Contents of the catalog}
\label{tabColumns}
\begin{footnotesize}
\begin{tabular}{p{0.04\linewidth}p{0.12\linewidth}p{0.7\linewidth}}
\tableline\tableline
Pos. & Label & Explanations \\ 
\tableline
1 & Name & The name or identifier of the star \\
2 & Bibcode & The ADS\tablenotemark{1} bibcode for the main source publication \\
3 & Type & Binnendijk type (``A" or ``W"; see Section \ref{secAWTypes} for details) \\
4 & ET & The eclipse type: $1$ for total, $0$ for partial eclipse \\
5 & $P$ & The orbital period of the binary in days \\
6 & $dPdt$ & The increase or decrease of the orbital period in seconds per year \\
7 & $P_c$ & The period of cyclical change of the orbital period in years \\
8 & BibcodeP & The ADS\tablenotemark{1} bibcode for the period study publication \\
9 & Solver & The software used to obtain the light curve solution (see Table \ref{tabSolvers} for details) \\
10 & $q$ & The mass ratio, $q = M_2/M_1$ (see Section \ref{secQT} for details) \\
11 & QT & Mass ratio type: spectroscopic (SP) or photometric (PH) \\
12 & $i$ & The orbital inclination of the binary in degrees \\
13 & $a$ & The orbital separation in solar radii \\
14 & Omega ($\Omega$) & The dimensionless Roche potential of the common envelope \\
15 & $f$ & The degree of overcontact or fillout degree (hereafter just fillout) \\
16 & $r_{1p}$ & Primary polar radius in units of orbital separation \\
17 & $r_{2p}$ & Secondary polar radius in units of orbital separation \\
18 & $T_1$ & Primary effective temperature in Kelvins \\
19 & $T_2$ & Secondary effective temperature in Kelvins \\
20 & $M_1$ & Primary mass in solar units \\
21 & $M_2$ & Secondary mass in solar units \\
22 & $R_1$ & Primary radius in solar units \\
23 & $R_2$ & Secondary radius in solar units \\
24 & $L_1$ & Primary luminosity in solar units \\
25 & $L_2$ & Secondary luminosity in solar units \\
26 & $L_3$ & Detection of uneclipsed (third) light ($1$ if detected, $0$ otherwise) \\
27 & $d$ & The estimated distance of the binary in parsecs \\
28 & Age ($\tau$) & The estimated age of the binary in Gyr \\
29 & Spots & Inclusion of spot(s) in the light curve solution ($1$ if included, $0$ otherwise) \\
\tableline
\end{tabular}
\tablecomments{This table is available in its entirety in the machine-readable format.
Only the column descriptions are shown here for guidance regarding its form and content.}
\tablenotetext{1}{The NASA's Astrophysics Data System Bibliographic Services (\textit{http://adsabs.harvard.edu/})}
\end{footnotesize}
\end{center}
\end{table}

\begin{table}
\begin{center}
\caption{Solver codes}
\label{tabSolvers}
\vspace{10pt}
\begin{tabular}{lll}
\tableline\tableline
Code & Solver & References \\ 
\tableline
WD & Various versions of Wilson \& Devinney software & 1, 2, 3 \\
L2 & Light2 & 4 \\
DC & Djurasevic code & 5, 6 \\
W3 & WUMA3 & 7 \\
BM & Various versions of the Binary Maker program & 8 \\
BS & BINSYN & 9 \\
PH & PHOEBE & 10, 11 \\
RO & ROCHE & 12 \\
LI & Linnell code & 13, 14 \\
\tableline
\end{tabular}
\tablerefs{1: \citet{wd1971}; 2: \citet{wd2008}; 3: \citet{wd2020ascl}; 4: \citet{light1993}; 
5: \citet{dc1992}; 6: \citet{dc1998}; 7: \citet{wuma1973}; 8: \textit{http://www.binarymaker.com/}; 9: \citet{binsyn1996}; 10: \citet{phoebe2005}; 11: \citet{phoebe2011ascl}; 12: \citet{roche2004}; 13: \citet{linnell1}; 13: \citet{linnell2}}
\end{center}
\end{table}


\subsection{Mass Ratio Transformation}
\label{secQT}

In this catalog, the mass ratio is defined as $q = M_2/M_1$, where the index $1$ denotes the more massive (primary), and the index $2$ the less massive (secondary) component so that $q$ is always less than 1. But many authors prefer to define the primary star as the one eclipsed in the deeper minimum regardless of mass, so that $q$ can be larger than 1 if the primary star defined this way is actually less massive than the companion, as is the case with W-type stars (see Section \ref{secAWTypes}). When we encounter a mass ratio larger than 1, we do the following transformations:

\begin{eqnarray}
\label{eqQT}
\notag q' &=& \frac{1}{q} \\
\Omega' &=& q' \Omega + \frac{1-q'}{2} \\
\notag Q_{1/2}' &=& Q_{2/1}
\end{eqnarray}

\noindent
where $Q$ can be any indexed quantity ($T, r, R, M, L$) and the transformed ($Q'$) quantities enter the catalog. This transformation was also applied in the rare cases\footnote{For example, \citet{DS2011} index the star eclipsed in the primary minimum with the subscript 1, and state that $M_2/M_1>1$ for W-type contact binaries, but report all the mass ratios as less than 1.} where the mass ratio reported in the source publication is less than 1, but $r_2>r_1$ and/or $M_2>M_1$.


\subsection{Missing Quantities}
\label{secMQ}

In the interest of maximizing the sample, we fill in the quantities missing from the source publications either by calculating them or by consulting other studies of the object in question.

In rare cases where the orbital period isn't reported in the light curve modeling paper, we use the one from the referenced period change study. In equally rare cases where the mass ratio isn't reported but the masses are, the missing mass ratio is calculated from the masses as $q=M_2/M_1$. The orbital separation is also calculated, if missing, from the period and masses whenever possible. When component luminosities are missing but either the absolute bolometric magnitudes or absolute radii are reported, we calculate the luminosities assuming the absolute bolometric magnitude and the effective temperature of the Sun are 4.74 and 5777 K respectively \citep{allen}.

Paired with the mass ratio, any one of the following quantities: $\Omega$, $f$, $r_{1, 2}$, which all provide a measure of relative star sizes, can be used to calculate all the others directly from the Roche model. The majority of solutions, especially those utilizing the WD code, list all of them, but there are still many cases where one or more are omitted, or where the authors report the geometric mean of the polar, side and back relative radii. We thus calculate $r_{1, 2}$ and $f$ from $\Omega$ whenever possible; or $r_{1, 2}$ and $\Omega$ from $f$, if $\Omega$ is not reported. If neither $\Omega$ nor $f$ are present, the solution is considered incomplete.

%
%

\section{Statistical analysis}
\label{secStats}

In the following sections, we examine the statistical properties of the sample. 

The cataloged data can be divided into three groups: categorical quantities (the eclipse and Binnendijk types, inclusion of spots and the third light etc.) stored either as boolean or text values; model parameters ($q$, $i$, $f$, $T_{1,2}$, $r_{1,2}$ and $\Omega$) and absolute parameters ($M_{1,2}$, $R_{1,2}$, $L_{1,2}$, $\tau$ and $d$), which have continuous numerical values. First, we look into the distributions of model parameters when objects are grouped according to the categorical quantities; and then into the overall distributions of model parameters and absolute parameters. 

Where appropriate, we fit the parameter distributions with gaussian mixtures according to the following model:

\begin{equation}
\label{eqGausFit}
g(x)=\sum_{j=1}^{m}\lambda_j\mathcal{N}(\mu_j, \sigma_j^2)
\end{equation}

\noindent
where $g$ is the fit to the observed quantity $x$, $m$ is the number of gaussian components, and $\mu$, $\sigma^2$ and $\lambda$ are the mean, variance and weight of each component. The fit is done using the R package \textit{mixtools}\footnote{https://www.R-project.org/}$^,$\footnote{http://www.jstatsoft.org/v32/i06/} \citep{mixtools} for analyzing finite mixture models. We ignore components with weights under 0.15.

\begin{figure}
\caption{The comparative distributions of selected quantities (columns) when the sample is divided according to selected categories (rows). See text for details.}
\label{figViolin}
\includegraphics[height=\textheight]{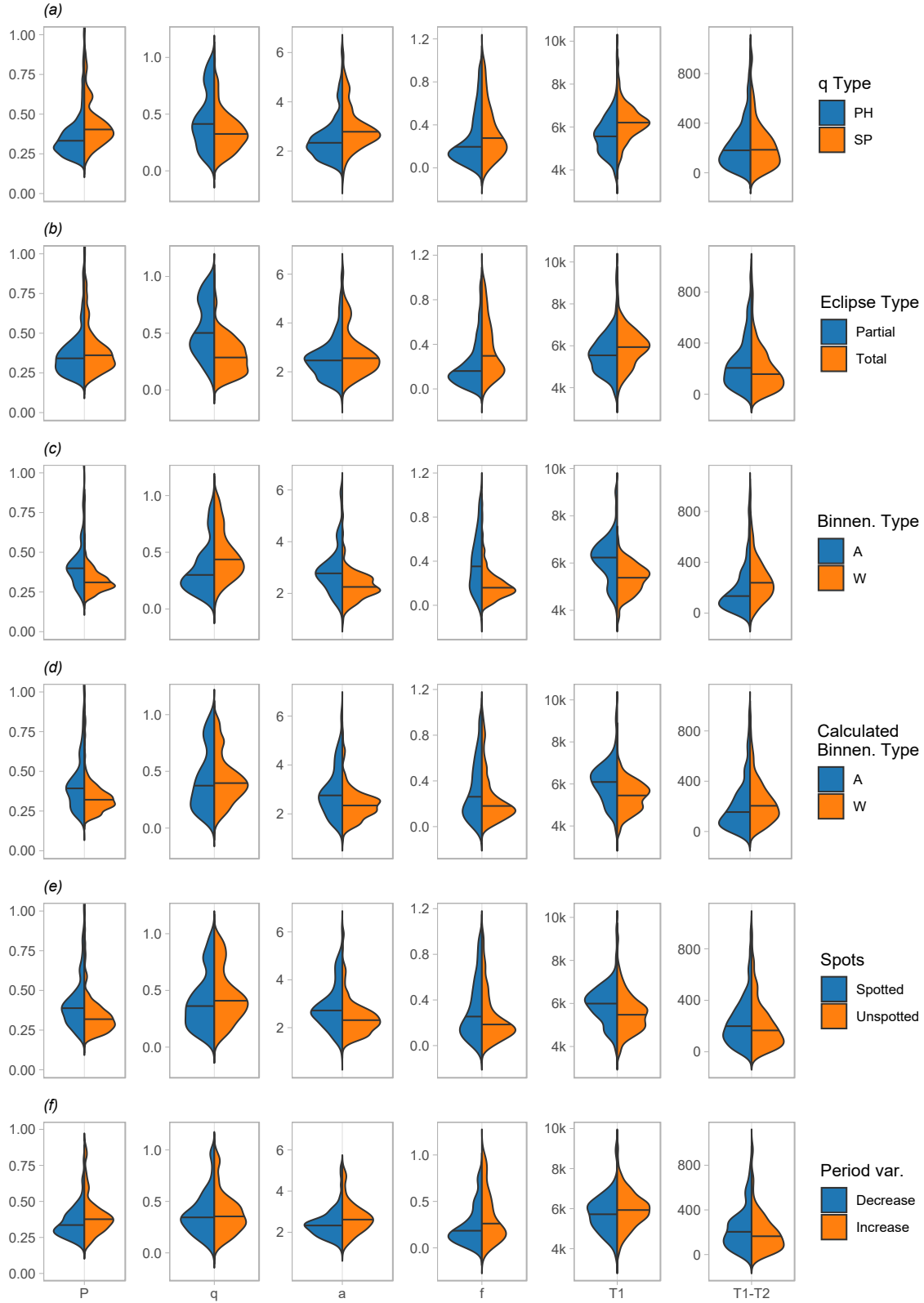}
\end{figure}

The comparative distributions of model parameters across different categories are shown in Fig. \ref{figViolin} using split violin plots\footnote{A ``violin plot" is the probability density derived from the histogram of the variable in question plotted vertically, with the ``split" between groups (e.g. the Binnendijk types) such that one group is plotted to the left, and the other to the right of the split and their similarity or difference is readily visible. The horizontal line in each distribution represents the median.}$^,$\footnote{In Fig. \ref{figViolin}, the distributions of the period and of the temperature difference are limited to 1 day and $|T_1-T_2| < 1000 K$, respectively, for clarity. The full distributions of these quantities are given in Sections \ref{secPdP} and \ref{secTRMLSA}.}. Summary statistics for numerical quantities are given in Table \ref{tabSummaryStats}. We proceed to discuss each category and distribution individually.

\begin{table}
\begin{center}
\caption{Summary statistics for numerical quantities in the catalog} 
\label{tabSummaryStats}
\begin{tabular}{lrrrrrrrrr}
\hline
Quantity  & min   & 5\%   & 10\%  & 25\%  & 50\% & 75\% & 90\% & 95\%  & max   \\ 
\hline
$P$       & 0.18  & 0.22  & 0.23  & 0.28  & 0.35 & 0.43 & 0.59 & 0.72  & 1.15  \\ 
$dPdt$    & -1.53 & -0.10 & -0.04 & -0.01 & 0.00 & 0.02 & 0.05 & 0.09  & 0.71  \\ 
$P_c$     & 1.77  & 3.57  & 5.44 & 10.54 & 19.70 & 38.18 & 57.16 & 76.63 & 139.00 \\ 
$q$       & 0.04  & 0.11  & 0.15  & 0.25  & 0.39 & 0.55 & 0.81 & 0.89  & 1.00  \\ 
$f$       & 0.00  & 0.01  & 0.05  & 0.11  & 0.20 & 0.43 & 0.66 & 0.80  & 1.00  \\ 
$i$       & 13.4  & 53.2  & 60.9  & 69.8  & 78.2 & 82.9 & 86.6 & 89.0  & 98.1  \\ 
$T_1$     & 3600  & 4403  & 4598  & 5063  & 5750 & 6328 & 6881 & 7194  & 9600  \\ 
$T_2$     & 3135  & 4428  & 4615  & 5119  & 5749 & 6250 & 6682 & 6976  & 9184  \\ 
$T_1-T_2$ & -2630 & -461  & -363  & -205  & 0    & 176  & 456  & 791   & 3083  \\ 
$r_1$     & 0.36  & 0.37  & 0.38  & 0.42  & 0.45 & 0.48 & 0.52 & 0.54  & 0.59  \\ 
$r_2$     & 0.16  & 0.21  & 0.23  & 0.26  & 0.29 & 0.32 & 0.35 & 0.36  & 0.41  \\
$R_1$     & 0.60  & 0.71  & 0.78  & 0.97  & 1.24 & 1.55 & 2.10 & 2.44  & 3.90  \\ 
$R_2$     & 0.19  & 0.47  & 0.51  & 0.61  & 0.77 & 0.91 & 1.18 & 1.35  & 2.20  \\ 
$M_1$     & 0.41  & 0.67  & 0.73  & 0.92  & 1.16 & 1.49 & 1.91 & 2.31  & 4.00  \\ 
$M_2$     & 0.08  & 0.16  & 0.18  & 0.27  & 0.38 & 0.57 & 0.82 & 0.97  & 1.88  \\ 
$L_1$     & 0.07  & 0.21  & 0.30  & 0.71  & 1.66 & 3.56 & 8.21 & 12.58 & 40.00 \\ 
$L_2$     & 0.02  & 0.10  & 0.16  & 0.33  & 0.64 & 0.98 & 1.78 & 3.04  & 11.27 \\ 
$a$       & 0.95  & 1.57  & 1.66  & 2.07  & 2.53 & 3.00 & 4.04 & 4.58  & 6.16  \\ 
$\tau$    & 0.51  & 1.89  & 2.41  & 3.42  & 4.45 & 6.54 & 8.76 & 10.85 & 13.51 \\ 
\hline
\end{tabular}
\end{center}
\end{table}


\subsection{Spectroscopic and Photometric Mass Ratios}
\label{secQcat}

It is interesting to divide the sample according to the method of mass ratio determination (photometric or spectroscopic, as defined in Section \ref{secOurCat}), as photometric mass ratios, typically obtained through a q-search\footnote{This usually involves creating a grid of models with fixed mass ratios and fitting their other parameters with light curve synthesis, then picking the mass ratio of the best-fitting model.} or direct fitting of synthetic light curves, are known to be less reliable than those measured from the radial velocities. 

The problem with photometric determination of the mass ratio is that the mass ratio and the orbital inclination affect the amplitude of the light curve in essentially the same way, and can therefore compensate for each other such that well-fitting models with a high and a low mass ratio can be found for the same light curve at different inclinations. However, this ambiguity is lifted when the system sports a total eclipse. With relative radii measurable from the eclipse duration, the inclination is no longer a confounding factor, and the mass ratio can be measured from photometric observations as reliably as from the radial velocities \citep{terrell2005}. In the next section, we compare the parameter distributions of totally and partially eclipsing systems, and find them remarkably distinct.

Our catalog contains \qSPCount \ objects with spectroscopic (``SP stars") and \qPHCount objects with photometric mass ratios (``PH stars"). The abundance of totally and partially eclipsing stars in both samples is similar. Fig. \ref{figViolin}a shows the comparative distributions of important numerical parameters in these two groups. The SP stars have longer periods, a single-peaked mass ratio distribution with a lower median and a smaller variance, and higher temperatures than the PH stars. These relations are similar to those that arise from dividing the sample into totally and partially eclipsing stars (Section \ref{secET} and Fig. \ref{figViolin}b).


\subsection{Eclipse Types}
\label{secET}

We divide the objects in the catalog into those with total (ET=1) and partial eclipses (ET=0, see Table \ref{tabColumns}). As we explained in the previous section, this distinction is important because in systems with total eclipses, the relative radii of components can be determined accurately from the eclipse duration, which in turn allows for a reliable estimate of the mass ratio.

The ability to infer the mass ratio from photometric observations alone is one of the most important aspects motivating the studies of contact binaries, but the type of eclipse is rarely stated explicitly by authors of such studies or included in tabular data detailing the solution. Perhaps that is the reason why this information can only be found in one of the previous catalogs mentioned in the Introduction \citep{PR2003}.

We determine the eclipse type from the source publications by visually examining the light curves. Where the eclipse type is not obvious and the authors make no comment on the presence or lack of totality, we catalog the object as having partial eclipses.

Of the \objCount objects in the catalog, \TECount are totally eclipsing and \PECount are partially eclipsing. The distribution of important parameters for these two groups are shown in Fig. \ref{figViolin}b. The totally eclipsing systems have larger fillouts and markedly smaller mass ratios than the partially eclipsing ones. The distribution of mass ratios for the first group has a median at $q=\TEmedianQ$ and a variance of \TEvarQ, while for the second group the median is at $q=\PEmedianQ$ with variance of \PEvarQ. As the mass ratio is determined from spectroscopy for only a relatively small number of objects (\qSPfracTE\% of the totally eclipsing and \qSPfracPE\% of the partially eclipsing subsample; see also Section \ref{secQcat}), the wide distribution in the second group reflects the poor reliability of photometric mass ratios derived for partially eclipsing systems.


\subsection{The ``A" and ``W" subtypes}
\label{secAWTypes}

According to the definition introduced by \citet{binnen70}, W UMa binaries are traditionally divided in two classes: the A-type and the W-type. In A-type systems, the deeper minimum is caused by the transit, and in the W-type systems, by the occultation of the less massive component. Due to the proportionality of the radius ratio and the mass ratio specific for contact systems (where the more massive star is always the larger of the pair), this distinction, formulated originally with regards to radial velocity curves, can be translated to the following criterion easily derived from a light curve solution: in A-type systems, the larger and more massive star is also the hotter one, and in W-type systems the smaller and less massive star is the hotter one. In A-type systems the total eclipse (if present) will be in the shallower, and in W-type systems, in the deeper minimum.

The differences in the effective temperatures of the components of W UMa systems are typically very small (see Section \ref{secTRMLSA}) so some authors have suggested that the apparently higher temperature of the secondary in W-type systems is due to the significant dark spot coverage of the primary, and not an intrinsic/evolutionary trait \citep[see e.g.][]{YE2005}. Others have tried to explain the W-type phenomenon in terms of stellar evolution within a contact system \citep[see e.g.][]{stepien2009, YD2013, YI2014, zhang2020a}. Conflicting theories emerged on whether the A- and W-types make an evolutionary sequence or have evolved from disparate initial conditions \citep[see e.g.][]{GS2008}.

Type determination can be ambiguous in partially eclipsing and spotted systems. Adding a cool spot to the hotter star or a hot spot to the cooler star may result in models that fit the observations equally well. Spots influence the estimated temperatures of both stars and may lead to the temperature ratio reversal, that might in turn result in different type assignments. On top of that, the eclipses are sometimes of equal depth.

Perhaps it is due to these problems that we find many works where the A/W subtype is assigned not according to the original definition, but according to secondary characteristics derived from early individual and statistical studies of W UMa stars \citep[see e.g.][]{MV1996}, which established expectations that the A-type stars should have higher temperatures (earlier spectral types), longer periods and smaller mass ratios than W-type stars. However, classifying W UMa stars following these expectations instead of the type definitions is a kind of a circular argument that has led, as we will see, to biased, artificially boosted differences between parameter distributions of the two groups.

Of the \AWCount objects in the catalog with type determination available in the source publications, \ATCount are of A type and \WTCount of W type. When type is instead calculated directly from the temperatures according to the \citet{binnen70} criterion:

\begin{equation}
\label{eqType}
\text{Type} = 
\begin{cases}
\text{A}, & \text{if } T_1 > T_2 \\
\text{W}, & \text{otherwise}
\end{cases}
\end{equation}

\noindent
we get \cATCount A types, \cWTCount W types, and \dAWCount objects where the type determined using Eq. \ref{eqType} differs from that in source publications. This is a significant fraction of our sample.

Fig. \ref{figViolin}c shows the distributions of key parameters for the A- and W-type stars with types adopted from source publications. The temperatures of primary components are higher by almost 1000 K in A- than in W-type systems, while the absolute temperature difference is lower by nearly a half. The periods are significantly longer ($P_{A,W}\approx \AmedP, \WmedP$), the fillouts larger ($f_{A,W}\approx \AmedF, \WmedF$) and mass ratios lower ($q_{A,W}\approx \AmedQ, \WmedQ$) in A- than in W-types.

However, when types are assigned using Eq. \ref{eqType}, the parameter distributions are notably less disparate between the two types (Fig. \ref{figViolin}d). While the temperatures of the primary components are still higher in A-type stars, the difference in median temperatures is around 500 K, which is not as dramatic as what we've seen previously. It is even more interesting to note that the distributions of the mass ratios and the absolute temperature differences no longer differ significantly. The periods and the fillouts still differ, but the differences are again smaller than before ($P_{A,W}\approx \cAmedP, \cWmedP$; and $f_{A,W} \approx \cAmedF, \cWmedF$ in A- and W-types, respectively).

Grouping the objects in the catalog according to Binnendijk types assigned using Eq. \ref{eqType} results in similar parameter distributions as grouping them according to the inclusion of spots in the model, which will be discussed in the next section (compare Fig. \ref{figViolin}d and Fig. \ref{figViolin}e). At first we thought this proved that the different properties of the Binnendijk types result from the presence of spots, as suggested by \citet{YE2005}. However, there seems to be no correlation between the spotted/unspotted groups and the A/W types, judging from Fig. \ref{figSTdP}a. The lack of variation between the number of A- or W-type stars among spotted and unspotted systems implies that these two groupings are independent.

In the rest of this work, we use the ``calculated'' types (assigned using Eq. \ref{eqType}).


\subsection{Spots}
\label{secSTypes}

Dark or bright spots are often included in the models of close binary stars, mostly to account for the so-called O'Connell effect: the asymmetry in the light curve where one maximum is lower than the other \citep{oconnell}\footnote{Some spot configurations (e.g. polar spots, or spots in the ``neck'' region between the components of a contact binary) may not cause discernible light curve asymmetry, but such cases are relatively rare.}. We catalog the inclusion (ST=1) or absence (ST=0) of spots in the light curve solution for each object (see Table \ref{tabColumns}). When multiple solutions are available in the source publication, with and without spots or with different kind of spots (dark or bright), we prefer the solution that fits the observations best (usually indicated by the $\chi^2$ metric). 

A statistical study of spot parameters (locations, temperatures and sizes) in a smaller sample of W UMa stars has been done by \citet{kozuma2019} with results that indicate different spot forming mechanisms in A- and W-type stars. In this work, we do not note the number of spots or the spot parameters, only the presence or absence of spots. As we have seen in the previous section, our data does not indicate a correlation between the types and the presence of spots. 

\begin{figure}
\caption{Top and middle: presence of spots and type of period change by Binnendijk type; bottom: type of period change by presence of spots. Abbreviations ``inc'', ``dec'' and ``cyc'' stand for increasing or decreasing period and cyclic period variation, respectively.}
\label{figSTdP}
\includegraphics[width=0.5\textwidth]{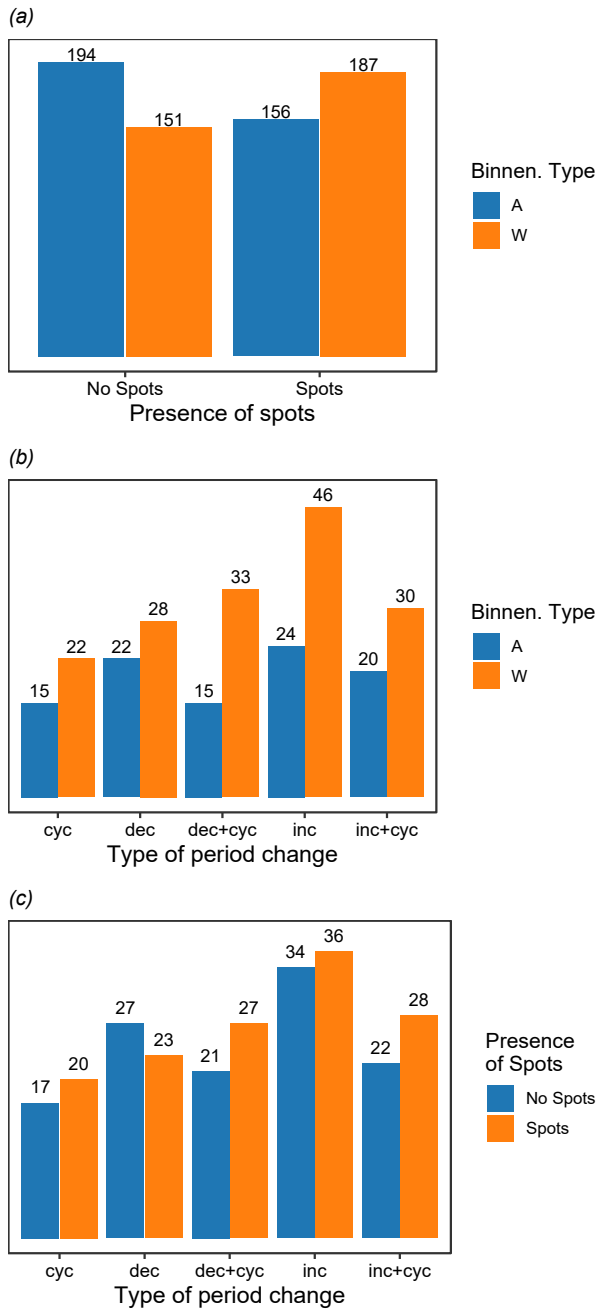}
\end{figure}

Since the presence of spots might be taken as an indicator of magnetic activity, and period changes can be interpreted in terms of magnetic cycles \citep{applegate}, we have also checked for correlations between the presence of spots and several period variation categories (see also Section \ref{secDPcat}): the detection or non-detection of period variability, the monotonic increase or decrease of the period, and the monotonic and cyclic variations. We find no notable correlations of spots with these quantities (Fig. \ref{figSTdP}\textit{c}).

Of the \objCount solutions in the catalog, \SPCount are spotted and \USCount are unspotted. The distributions of model parameters for these two groups are shown in Fig. \ref{figViolin}e. They are similar to those of A/W type stars, respectively. Spotted solutions exhibit higher temperatures, longer orbital periods and larger fillouts.

\subsection{The ``B" and ``H" subtypes}
\label{secBH}

Apart from the A/W subtypes, W UMa stars are sometimes also classified as belonging to the ``B" and ``H" subtypes in literature. 

The B subtype designates contact binaries in poor thermal contact, that is, such where the difference in the effective temperatures of the components is significant. The designation B comes from the similarity of their light curves to those of $\beta$ Lyr stars. It was introduced by \citet{lucy79}, who studied three such W UMa stars (AK Her, W Crv and RW PsA) and found, other than a temperature difference that was considered large at the time but would not be outstanding in the current sample (370 K, 650 K and 275 K, respectively), that anomalous values of albedos and gravity darkening were required to fit the observations (which was disproved later, using models that allow the inclusion of spots; \citealp[see][]{caliskan2014}). In more recent works \citep[e.g.][]{CK2004}, B-type stars are defined by a temperature difference in excess of 1000 K.

The distribution of temperature differences across the entire sample, presented in Section \ref{secTRMLSA}, is centered near 0 K and is about 250 K wide, with 95\% of the stars having $\Delta T \lessapprox 800 K$. This confirms that the limit of 1000 K for the B-type stars is appropriate. According to this definition, there are only \BCount B-type stars in our catalog. They tend to have longer periods, larger separations, higher primary temperatures and lower fillouts than the rest of the sample.

The H subtype designates contact binaries with a large mass ratio. It was introduced by \citet{CK2004}, who found that systems with mass ratios larger than $q = 0.72$ form a distinct group in terms of energy transfer. \citet{MV1996} also found a separation between systems of low and high mass ratios in the distributions of period, mass ratio and fillout. The joint sample of 26 H type stars from these two papers contains 18 objects that are also included in our catalog; among these, 10 are partially eclipsing with a photometric mass ratio, and we have seen in Sections \ref{secQcat} and \ref{secET} that such mass ratios tend toward large values and should be considered tentative at best. We speculate that the grouping might not have been noticed at all if stars with unreliable mass ratios had been omitted.

Of the \objCount stars in our catalog, \HCount belong to the H type according to the criterion of \citet{CK2004}. The H stars have lower temperatures, shorter periods and smaller fillouts than non-H stars. However, only \HCountRel H stars have reliable (spectroscopic, or photometric with total eclipses) mass ratios.

Unlike the assignment of the A/W types, which is something of a de facto standard in modern studies of W UMa stars, the B and H types are used rarely. Since the sample is easily divided in B and non-B (or H and non-H) objects with a hard limit on a single parameter, these classifications are not included in the catalog.


\subsection{Third Light}
\label{secL3}

Adding the third (uneclipsed) light to the model is sometimes necessary for achieving a good fit to the observations. We indicate the presence (L3=1) or the absence (L3=0) of third light for all the objects in the catalog. Among the previous catalogs, only the one presented by \citet{CK2004} also carried this information.

Third light is a poorly constrained parameter that affects the amplitude of the light curve. As such, it may greatly influence the other parameters of the model, especially the inclination and the photometric mass ratio. Usually it is not adjusted unless it is impossible to obtain a decent fit to the observations otherwise. Authors often offer both solutions: one with and the other without the third light and as a rule, the former has a formally better fit and will therefore be the one we include in the catalog (see Section \ref{secOurCat}). Because of this, the frequency of non-zero third light contribution reported here likely overestimates the occurrence rate of actual light pollution by an unresolved source.

Of the \objCount objects in our sample, \HL3Count have a non-zero, and \NL3Count zero third light contribution. Distributions of selected parameters with respect to the presence of the third light do not differ significantly. Solutions with third light tend to have larger and more uniformly distributed fillouts and slightly lower mass ratios than the rest of the sample.


\subsection{Orbital Period Variation as a Categorical Quantity}
\label{secDPcat}

Secular period variations are expected in all, and observed in hundreds of contact binaries \citep[see e.g.][]{kubiak, li2018}. The period may decrease or increase monotonically due to the mass transfer or the changes in separation induced by angular momentum loss. It may also vary periodically, indicating that the binary is a part of a multiple system, or as a result of cyclic magnetic activity \citep{applegate}. These mechanisms are discussed widely in the literature and summarized in e.g. \citet{liu2018}, who propose that the thickness of the common envelope also plays an important role.

First we look into the period decrease and increase, with or without the cyclic change, as categorical variables that we try to correlate with the A/W types and other parameters; then, we look at the distribution of the measurements.

Of the \dPCount objects in the catalog with a period change measurement, \PincCount have monotonically increasing, \PdecCount monotonically decreasing, and \PcCount a period with cyclic variation; \dPcCount objects show both the monotonic and cyclic period changes. Fig. \ref{figSTdP}b shows the objects with period change measurements, grouped by the A/W type. For all types of period change, the W-type stars are more numerous, but there's no evidence that either type prefers any single kind of variation.

From Fig. \ref{figViolin}f, we see that the type of period change doesn't influence the distributions of selected quantities, with the exception of fillout, which is notably larger and more widely distributed in stars with increasing periods. A similar trend, though less pronounced, is visible with the period.


\subsection{Orbital Period and Period Variations as Continuous Quantities}
\label{secPdP}

\begin{figure}
\caption{Top: period distribution of contact binaries in our catalog, compared with LAMOST and CRTS data. Bottom left: monotonic period change distribution for the objects in our catalog and OGLE contact binaries. Bottom right: cyclic period change distribution. The solid (dotted) lines represent the sum (individual components) of the gaussian fit to the catalog data.}
\label{figDistroPdP}
\includegraphics[width=\textwidth]{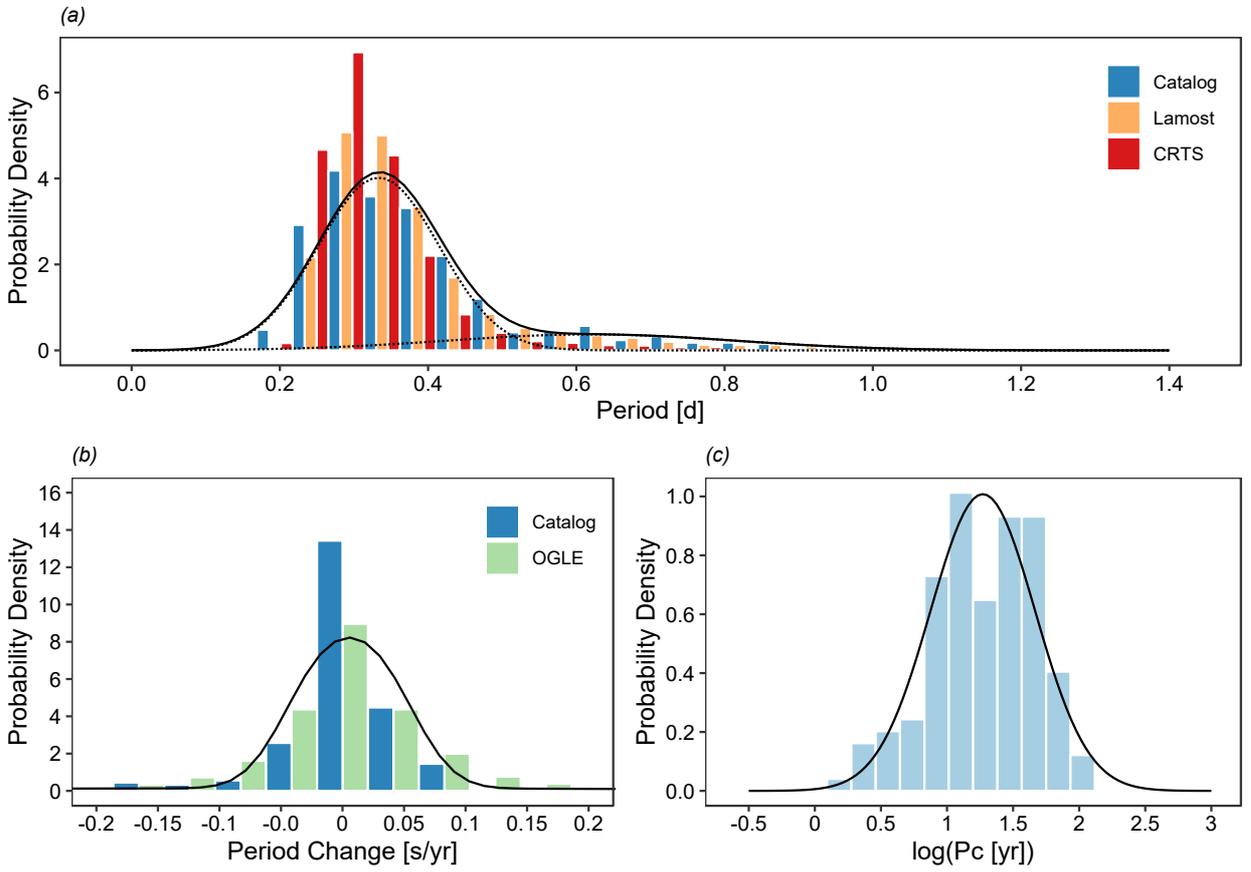}
\end{figure}

The distributions of the orbital periods and the monotonic and cyclic period variations of the cataloged objects are shown in Fig. \ref{figDistroPdP}. 

We compare the period distribution of our sample with the the much more numerous samples from LAMOST (\lamostCount objects) and CRTS (\crtsCount objects). Increasing the sample size seems to enhance the sharpness of the distribution peak near the value of 0.3 days. The objects in our catalog tend towards shorter periods, including \objPus ultra-short-period objects with periods under the 0.22 days cutoff \citep{rucinski1992, zhang2020b}. 

The parameters of the two-component gaussian fit (Eq. \ref{eqGausFit}) to the catalog data are: $\mu_1=\Pamu$ and $\sigma^2_1=\Pasigma\ d$ ($\lambda_1=\Palambda$), and $\mu_2=\Pbmu$ and $\sigma^2_2=\Pbsigma\ d$ ($\lambda_2=\Pblambda$). Half of the objects in the catalog have periods between \Ploq and \Pupq $d$, with the median at \Pmed $d$. The system with the shortest period is \objPmin ($P=\Pmin\ d$).

The histogram of the monotonic orbital period change in cataloged objects conforms to a single-peak normal distribution with the mean at $\mu=\dPmu\ s/y$ and variance of $\sigma^2=\dPsigma\ s/y$. Half of the sample falls between \dPloq and \dPupq $s/y$, with the median at \dPmed $s/y$. The system with the fastest period change is \objdPmax ($dP/dt=\objdPmaxVal\ s/yr$). Also shown in Fig. \ref{figDistroPdP}\textit{b} are the period change measurements for a sample of 500 contact binaries from OGLE \citep{kubiak}. Their distribution is in good agreement with our findings.

The histogram of the periods of the cyclic orbital period variations of cataloged objects can also be approximated by a single-peak log-normal distribution with $\mu=\PCmu$ and $\sigma^2_1=\PCsigma$. Half of the sample has variation periods between \PCloq and \PCupq $y$, with the median at \PCmed $y$. The system with the shortest period of cyclic change is \objPCmin ($Pc=\PCmin\ y$).

\subsection{Mass Ratio, Fillout and Orbital Inclination}
\label{secQFI}

\begin{figure}
\caption{The distributions of mass ratios (top), fillouts (bottom left) and orbital inclinations (bottom right) for contact binaries in our catalog. The solid (dotted) lines represent the sum (individual components) of the gaussian mixture fit.}
\label{figDistroQFI}
\includegraphics[width=\textwidth]{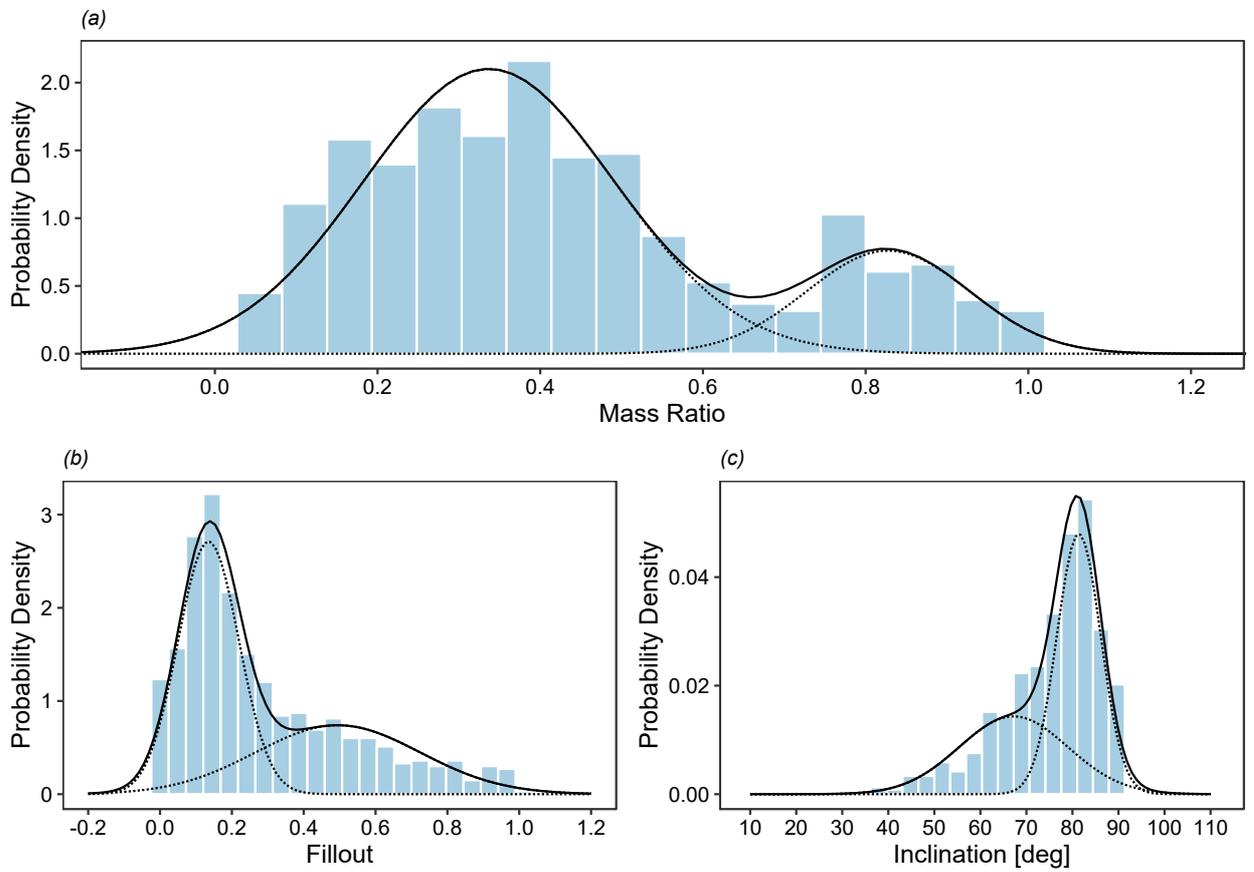}
\end{figure}

The mass ratio distribution of the objects in the catalog is shown in Fig. \ref{figDistroQFI}a with the two-component gaussian fit (Eq. \ref{eqGausFit}). The parameters of the components are: $\mu_1=\Qamu$, $\sigma^2_1=\Qasigma$, $\lambda_1=\Qalambda$ and $\mu_2=\Qbmu$, $\sigma^2_2=\Qbsigma$, $\lambda_2=\Qblambda$. Half of the stars have mass ratios between \Qloq and \Qupq, with the the median at \Qmed. The system with the smallest mass ratio is \objQmin ($q=\Qmin$), well below the theoretical lower limit \citep{arbo2007, arbo2009}.

The stars in the catalog form two distinct groups: one centered around $q=0.33$ and another, shifted toward greater values and covering a greater range. These groups roughly correspond to systems that manifest total and partial eclipses, respectively, as can be seen in Fig. \ref{figViolin}. A large majority of totally eclipsing systems have mass ratios distributed uniformly in the range from about 0.1 to 0.5, while partially eclipsing stars have a double-peaked distribution covering the entire range from 0.1 to 1.0. As discussed in Sections \ref{secQcat} and \ref{secET}, the mass ratio is poorly constrained by the light curves of partially eclipsing systems, resulting in a smeared distribution. Since only \qSPCount \ out of \objCount systems in the catalog have a spectroscopically determined mass ratio, the distribution of this quantity is dominated by objects with photometric mass ratios.

Fig. \ref{figDistroQFI}b shows the distribution of the fillouts. The parameters of the two-component gaussian fit are: $\mu_1=\Famu$, $\sigma^2_1=\Fasigma$, $\lambda_1=\Falambda$ and $\mu_2=\Fbmu$, $\sigma^2_2=\Fbsigma$, $\lambda_2=\Fblambda$. Again, these two components roughly correspond to partially and totally eclipsing stars, as can be seen in Fig. \ref{figViolin}. Half of the objects in the catalog have fillouts between \Floq and \Fupq, with the median at \Fmed. There are \fZeroCount stars with fillout equal to zero, corresponding to binaries in exact contact\footnote{\FZObjs}. The star with the largest fillout is \objFmax ($f=\Fmax$).

The distribution of the orbital inclinations is shown in Fig. \ref{figDistroQFI}c. The parameters of the two-component gaussian fit are: $\mu_1=\Iamu^{\circ}$ and $\sigma^2_1=\Iasigma^{\circ}$ ($\lambda_1=\Ialambda$), and $\mu_2=\Ibmu^{\circ}$ and $\sigma^2_2=\Ibsigma^{\circ}$ ($\lambda_2=\Iblambda$). Here as well, the two components roughly correspond to partially and totally eclipsing stars (Fig. \ref{figViolin}). Half of the stars in our sample have inclinations between $\Iloq^{\circ}$ and $\Iupq^{\circ}$, with the median at $\Imed^{\circ}$. The tendency of catalog objects toward high inclinations is a selection effect caused by the nearly sinusoidal shape of the light curves of contact binaries: at lower inclinations, the eclipses, which are crucial for successful light curve modeling, become difficult to distinguish from the ellipsoidal variation.

\subsection{Absolute Parameters}
\label{secTRMLSA}

\begin{figure}
\caption{The distributions of primary temperatures, temperature differences, relative radii, primary absolute radii, masses and luminosities in solar units, orbital separations and ages of the objects in our catalog, compared, where possible, with LAMOST, CRTS and Gaia data. The solid (dotted) lines represent the sum (components) of the gaussian fits to catalog data.}
\label{figDistroTRMLSA}
\includegraphics[height=0.9\textheight]{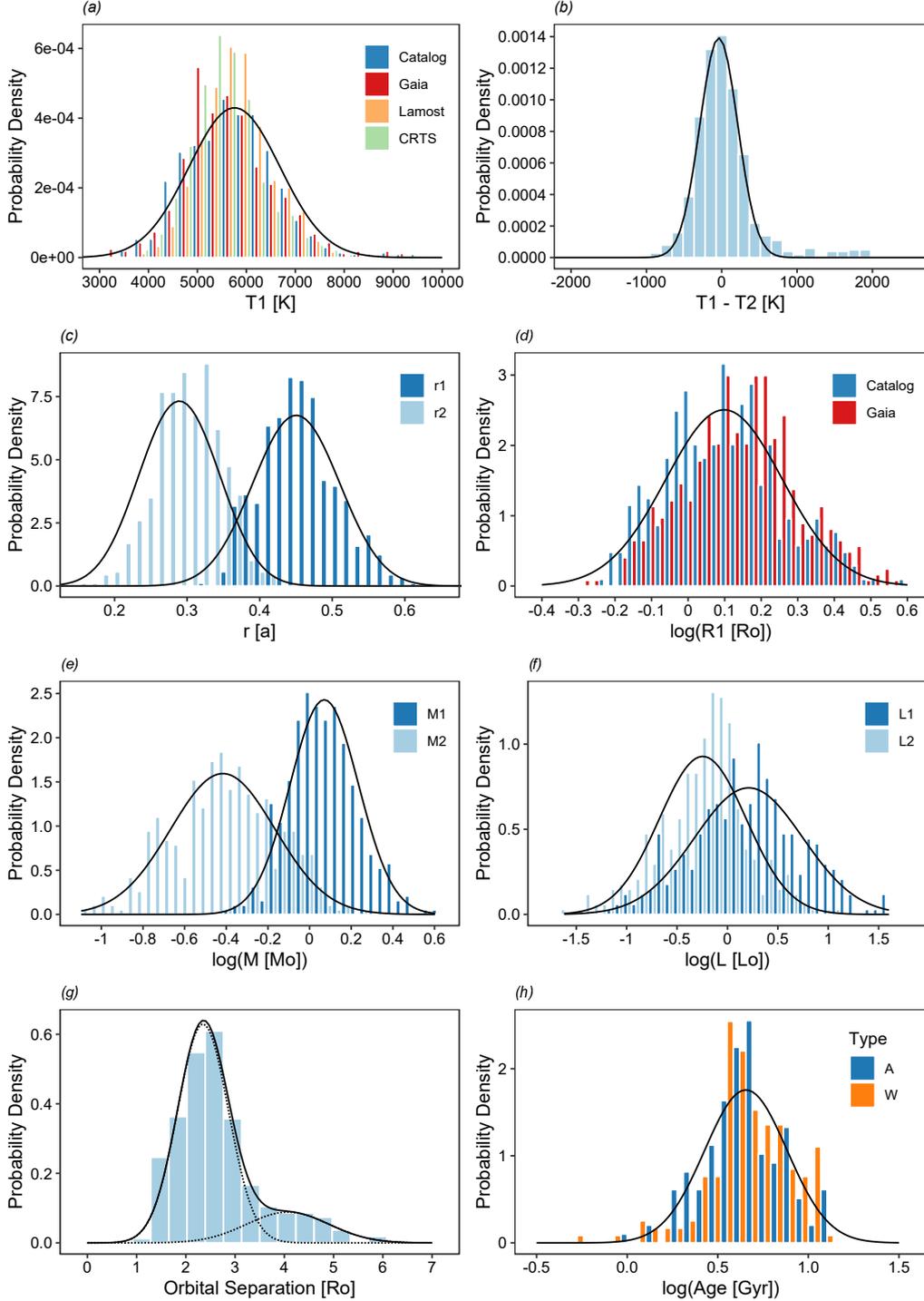}
\end{figure}

The primary temperature distribution of the objects in the catalog is shown in Fig. \ref{figDistroTRMLSA}a. It can be described by a gaussian fit centered at $\mu = \Tmu$ K with the standard deviation of $\sigma = \Tsigma$ K. Half of the stars have primary temperatures between \Tloq K and \Tupq K. The system with the lowest primary temperature is \objTmin ($T_1=\Tmin$ K), and with the highest, \objTmax ($T_1=\Tmax$ K).

Secondary temperatures have the same distribution as the primary temperatures, which is to be expected in contact binaries. The distribution of temperature differences between the components ($\delta T=T1-T2$) is shown in Fig. \ref{figDistroTRMLSA}b. It is best approximated with a single-peak gaussian of mean $\mu=\TDamu$ K and variance $\sigma^2=\TDasigma$ K. For half of the stars in our sample, the temperature difference is between \TDloq and \TDupq K, with the median at \TDmed K. The distribution tail comprising values in excess of about 750 K represents the B-type contact binaries. 

Fig. \ref{figDistroTRMLSA}c shows the distributions of primary and secondary polar radii in units of orbital separation (the so-called relative radii) together with gaussian fits. Primary (secondary) relative radii are distributed around $r_1=\rimu$ ($r_2=\rzmu$) with variance of $\sigma^2_{r_1}=\risigma$ ($\sigma^2_{r_2}=\rzsigma$). A smaller sample of \RCount objects has estimates of radii in solar units (absolute radii). The distribution of primary absolute radii is compared in Fig. \ref{figDistroTRMLSA}d with the distribution of Gaia radius measurements. The distributions are fairly similar and centered slightly above the solar radius. The Gaia distribution is shifted toward larger values compared to the catalog data as a result of calculating the radius from the total system luminosity \citep{gaia1, gaia2}.

There are \mCount stars with estimates of component masses and \lCount with estimates of component luminosities in solar units in the catalog. The distributions of these quantities are shown in Figs. \ref{figDistroTRMLSA}e and \ref{figDistroTRMLSA}f. Primary (secondary) masses and luminosities are distributed around $\log M_1=\mimu$ ($\log M_2=\mzmu$) and $\log L_1=\limu$ ($\log L_2=\lzmu$) with variance of $\sigma^2_{M_1}=\misigma$ ($\sigma^2_{M_2}=\mzsigma$) and $\sigma^2_{L_1}=\lisigma$ ($\sigma^2_{L_2}=\lzsigma$). The most massive object in the catalog is \objMmax, with the total mass of $\Mmax M_{\odot}$, and the least massive \objMmin, with the total mass of $\Mmin M_{\odot}$.

The estimates of orbital separation are available for \ACount stars in the catalog. Their distribution, shown in Fig. \ref{figDistroTRMLSA}g, can be described by a two-component gaussian fit with the following parameters: $\mu_1=\Aamu$ and $\sigma^2_1=\Aasigma\ \rm R_{\odot}$ ($\lambda_1=\Aalambda$), and $\mu_2=\Abmu$ and $\sigma^2_2=\Absigma\ \rm R_{\odot}$ ($\lambda_2=\Ablambda$). Half of the stars have separations between \Aloq and \Aupq $\rm R_{\odot}$. The object with the smallest separation is \objAmin ($a=\Amin\ \rm R_{\odot}$), and with the largest, \objAmax ($a=\Amax\ \rm R_{\odot}$).

\subsection{Ages}
\label{secAge}

We estimate the ages of all the stars in the catalog using the methods developed by \citet{YD2013} and \citet{YI2014}, which rely on empirical relations derived by comparing evolutionary models with a sample of about 100 W UMa stars whose physical parameters have been determined robustly through combined spectroscopic and photometric investigations\footnote{The sample used by \citet{YD2013} and \citet{YI2014} is a subset of what we call the ``old sample" in the Introduction. Since the parameter values in the present work come from the latest publications (see Section \ref{secOurCat}), they might differ slightly from the values quoted in those papers; consequently, so might the ages.}. The procedure of age calculation as adopted for this work is outlined in Appendix \ref{apxAgeCalc}.

We were able to obtain plausible age estimates for \GCount objects with mass and luminosity measurements. Due to the approximate nature of the method, some estimates resulted in ages longer than the age of the universe, and these were omitted from the catalog. The rest of the sample is in overall agreement with the kinematics studies of \citet{bilir2005} and \citet{rucinski2013}.

The distribution of the ages is shown in Fig. \ref{figDistroTRMLSA}h. It can be described by a log-normal function centered at \Gmu Gyr with the standard deviation of \Gsigma Gyr. Half of the objects with the age estimate are in the range from \Gloq to \Gupq Gyr. The youngest system in this sample is \objGmin ($\tau=\objGminG$ Gyr).

%
%

\section{Absolute parameters as functions of mass ratio and period}
\label{secRel}

In this section we examine how the absolute parameters depend on the mass ratio and the period, as well as a few other interesting relations between the cataloged quantities. 

Based on the findings presented in Sections \ref{secQcat}, \ref{secET} and \ref{secQFI}, we group the objects according to the reliability of the mass ratio determination into the reliable (with either a spectroscopic mass ratio or a total eclipse) and the unreliable sample (with partial eclipses and no spectroscopic support).

It is important to note, however, that the reliability of the mass ratio does not necessarily imply the reliability of absolute parameters. Without radial velocities to constrain the size of the orbit, the stellar masses and absolute radii cannot be determined from the light curves directly. The usual method to estimate these quantities is to assume that the primary component is on the main sequence, and adopt its mass from its spectral type; the mass of the companion can then be calculated from the mass ratio, and the orbital separation from Kepler's third law. Other indirect methods can be employed too \citep[see e.g.][]{GS2008}. Additional uncertainty in the estimation of luminosities comes from the determination of component temperatures, which is usually based on the spectral type or the color index of the binary and again cannot be made from the light curves directly. The absolute parameters for most of the stars in the catalog thus depend on external assumptions that differ from one study to the next and result in inhomogeneity which is the main source of the large scatter seen in the statistical relations presented below.

Fig. \ref{figRelQ}a shows the dependence of the primary temperature on the mass ratio. The objects are divided by type (represented by different colors) and the reliability of the mass ratio (represented by different shapes). The primary temperatures of systems with mass ratios up to about 0.5 follow a downward trend (solid line) given by Eq. \ref{eqRelQT}. The A-type stars in this part of the diagram can be seen to have higher temperatures and smaller mass ratios compared to the W-type stars, but there is no exclusive grouping.

\begin{equation}
\label{eqRelQT}
    T_1= -(2752 \pm 320) P + (6745 \pm 100)
\end{equation}

\noindent Above the mass ratio of 0.5, there are hardly any reliable systems and the objects in this region show no dependence of the temperature from the mass ratio (dashed line in Fig. \ref{figRelQ}a).

Panels \textit{b, c} and \textit{d} of Fig. \ref{figRelQ} show the dependence of absolute parameters for the primary and secondary components of cataloged stars on the mass ratio. The masses, radii and luminosities of the primary components follow downward, and those of the secondary components, upward trends. Only the reliable sample (see Table \ref{tabSamples}) is used for fitting the following relations:

\begin{eqnarray}
    \label{eqRelM1Q}
    M_1 = (1.39 \pm 0.57) q^2 - (2.01 \pm 0.49) q + (1.83 \pm 0.09) \\ 
    M_2 = -(0.05 \pm 0.20) q^2 + (1.06 \pm 0.17) q + (0.08 \pm 0.03) \\
    R_1 = (2.07 \pm 0.57) q^2 - (2.83 \pm 0.49) q + (2.06 \pm 0.09) \\ 
    R_2 = -(0.03 \pm 0.33) q^2 + (0.49 \pm 0.28) q + (0.65 \pm 0.05) \\
    L_1 = (4.08 \pm 1.72) q^2 - (6.57 \pm 1.49) q + (3.53 \pm 0.28) \\ 
    \label{eqRelL2Q}
    L_2 = -(0.54 \pm 0.60) q^2 + (0.84 \pm 0.53) q + (0.41 \pm 0.10)
\end{eqnarray}

Fig. \ref{figRelQ} also shows the orbital period (panel \textit{e}) and its monotonic variation (panel \textit{f}) as functions of the mass ratio. Like with the primary temperature (\ref{figRelQ}\textit{a}), the periods in the reliable sample follow a downward trend up to $q\approx 0.5$, while the unreliable sample doesn't seem to vary with the mass ratio at all. From Fig. \ref{figRelQ}f, we can also see that the speed and direction of the period change do not depend in any notable way on the mass ratio or on the type of the system.

\begin{figure}
\caption{The dependence of the primary temperature, absolute parameters, period and period variability on the mass ratio. The labels ``R" and ``U" refer to the reliable and unreliable sample, respectively. Lines represent the fits through reliable data, with shading indicative of uncertainty. See text for more details.}
\label{figRelQ}
\includegraphics[width=\textwidth]{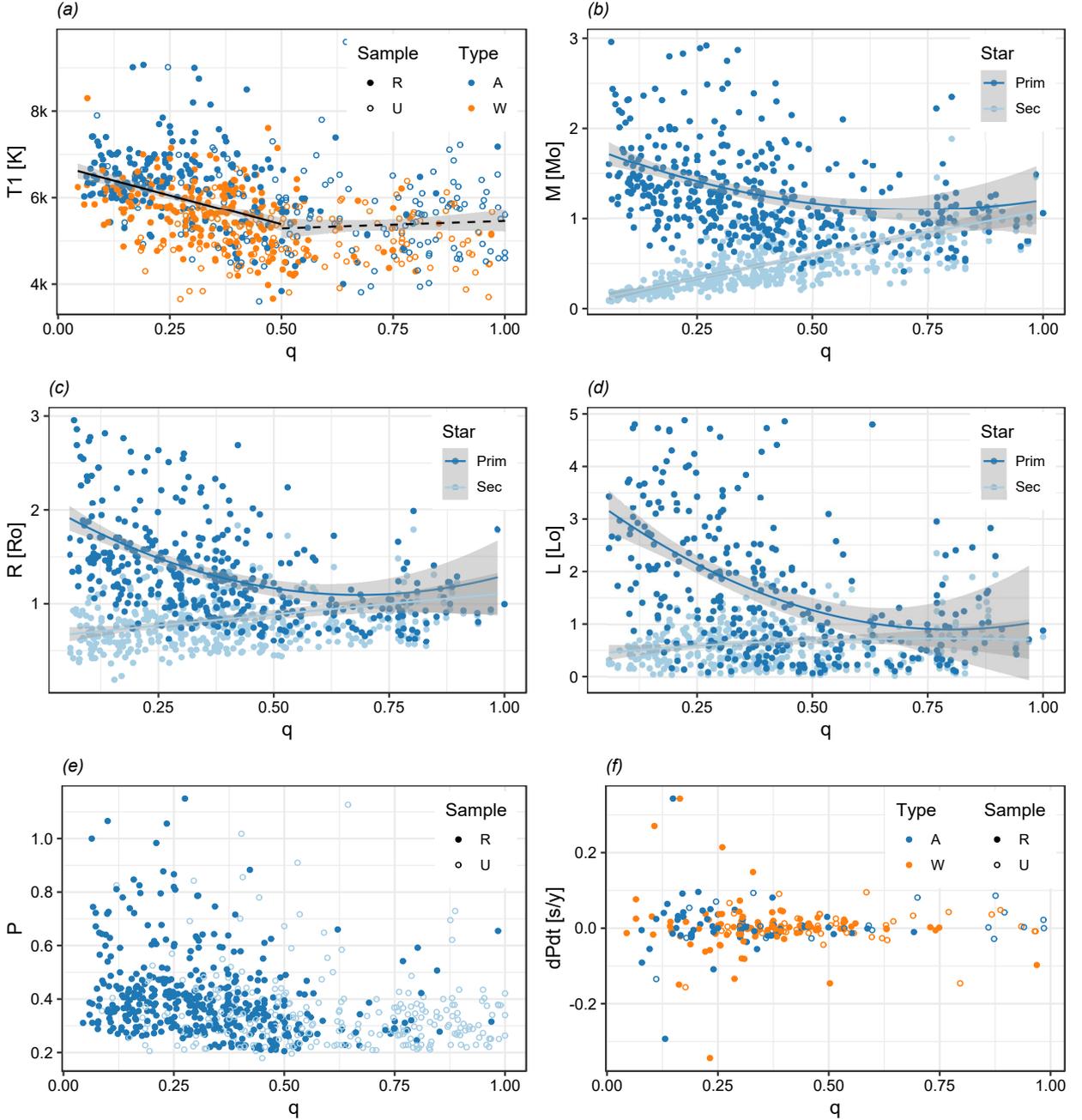}
\end{figure}

\begin{figure}
\caption{The dependence of the primary temperature and absolute parameters on the orbital period. The gray crosses in panel \textit{a} represent the CRTS sample.}
\label{figRelP}
\includegraphics[width=\textwidth]{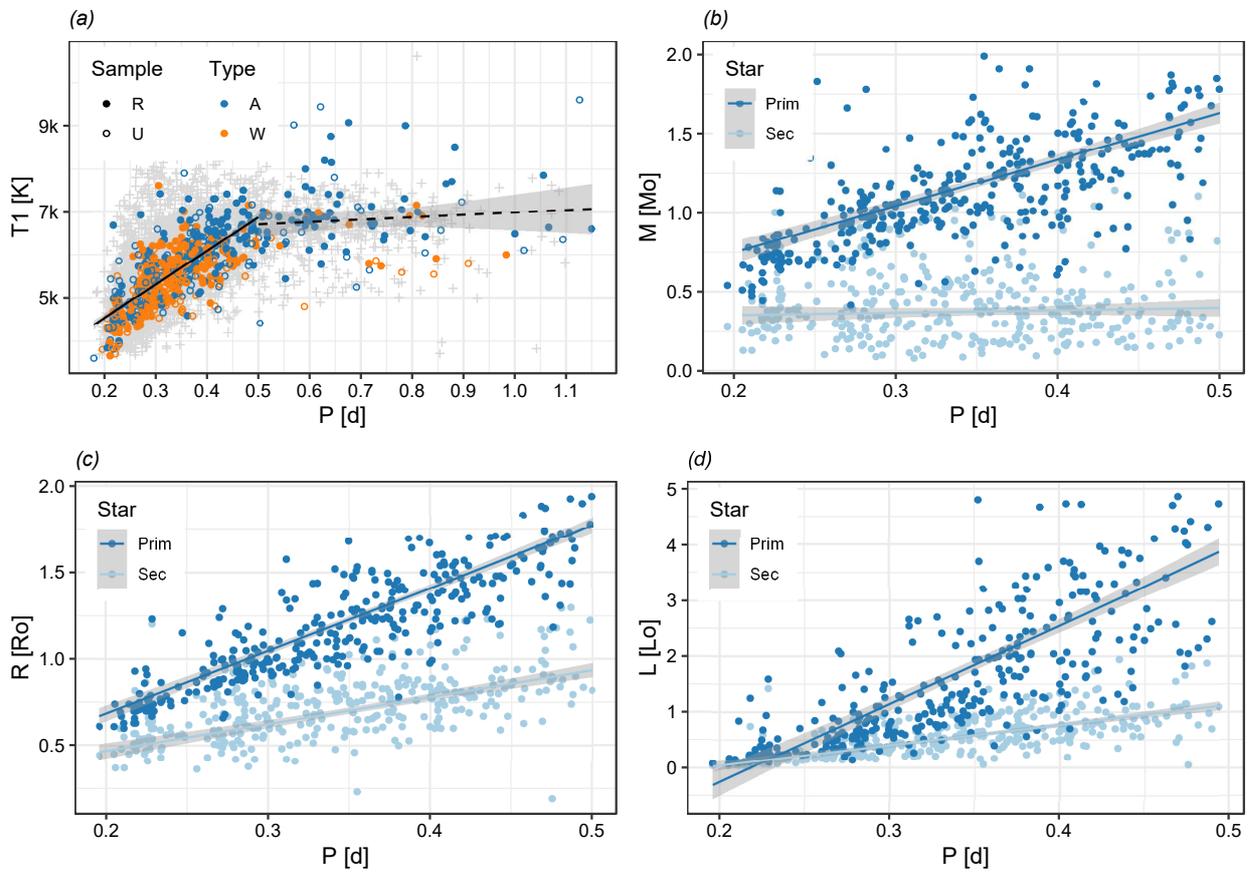}
\end{figure}

In Fig. \ref{figRelP}, we plot the dependence of the primary temperature (panel \textit{a}), component masses (\textit{b}), radii (\textit{c}) and luminosities (\textit{d}) on the orbital period. Panel \textit{a} also shows the much more numerous sample of W UMa stars from CRTS as gray crosses. Like in Fig. \ref{figRelQ}a, different colors represent the types, and different shapes the reliability of the mass ratio. An upward trend in primary temperatures given by Eq. \ref{eqRelPT} is seen among the objects with periods up to about 0.5 days (solid line). The trend disappears for objects with longer periods (dashed line). \citet{asassn} noted a similar break in the period-temperature diagram of a large sample of ASAS-SN contact binaries and attributed it to the Kraft transition from slow to fast rotation that occurs between 6200 and 6700 K \citep{kraft}. This is also the range of temperatures separating binaries with convective and radiative envelopes \citep{kippenhahn}. 

\begin{equation}
\label{eqRelPT}
    T_1= (7780 \pm 307) P + (2977 \pm 105)
\end{equation}

While it might be tempting to equate contact binaries with radiative envelopes located above the Kraft transition with the A-type stars, our data rather suggests that such objects shouldn't be labeled as W UMa stars to begin with. Both the A- and the W-type objects are well-represented in the region of the period-temperature diagram under 0.5 days and below 7000 K, and although under-represented, W-type stars with reliable parameters are still seen outside of it. A more limiting definition of W UMa stars, such that would exclude stars with radiative envelopes, would better describe the observed groupings and tendencies.

The masses, radii and luminosities for the primary and secondary components of cataloged stars as functions of orbital period, shown in panels \textit{b, c} and \textit{d} of Fig. \ref{figRelP}, can be approximated by the following relations:

\begin{eqnarray}
    \label{eqRelMRLP}
    M_1 = (2.94 \pm 0.21) P + (0.16 \pm 0.08) \\ 
    M_2 = (0.15 \pm 0.17) P + (0.32 \pm 0.06) \\
    R_1 = (3.62 \pm 0.13) P - (0.04 \pm 0.05) \\ 
    R_2 = (1.56 \pm 0.13) P + (0.16 \pm 0.05) \\
    L_1 = (13.98 \pm 0.75) P - (3.04 \pm 0.27) \\ 
    L_2 = (3.66 \pm 0.26) P - (0.69 \pm 0.09)
\end{eqnarray}

As in equations \ref{eqRelM1Q}--\ref{eqRelL2Q}, only the reliable sample is used for fitting (see Table \ref{tabSamples} for details).

\begin{figure}
\caption{The dependence of the temperature difference on the fillout (left) and the primary temperature (right).}
\label{figRelDT}
\includegraphics[width=\textwidth]{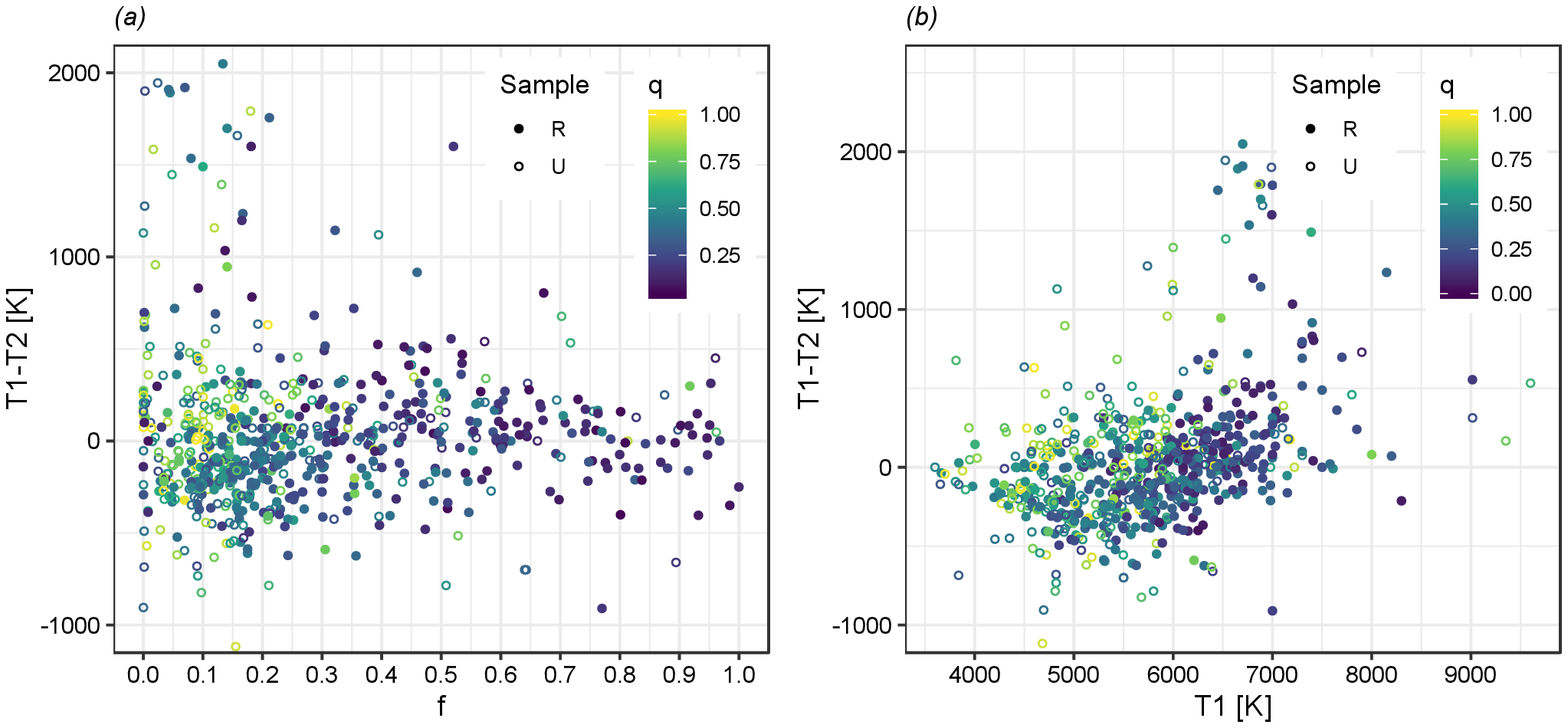}
\end{figure}

Fig. \ref{figRelDT} shows the temperature difference, $\delta T = T_1-T_2$, as a function of fillout (panel \textit{a}) and the primary temperature (panel \textit{b}). The correlation between $\delta T$ and the fillout is weak. The objects with greatest $\delta T$ do pile up in the fillout range from zero to about 20\%, but the majority of these objects come from the unreliable sample. Otherwise, the $f$ vs $\delta T$ diagram reflects the distribution of temperature differences without any discernible trend and objects with $\delta T$ in excess of 500 K abound for fillouts up to 50\%. The data in our catalog doesn't bear the intuition that shallow (deep) contact should lead to greater (smaller) temperature differences.

When $\delta T$ is instead plotted against the primary temperature, the majority of objects forms a tight group with $|\delta T| < 500 K$ and $4500 K < T_1 < 7000 K$. Only a sparse scattering of (mostly unreliable) systems is found outside of this range. We have seen the same grouping in the $T_1$ vs $q$ and $T_1$ vs $P$ diagrams (Figs. \ref{figRelQ}\textit{a} and \ref{figRelP}\textit{a}).

\begin{figure}
\caption{Top: the HR diagram of the stars in the catalog with MIST isochrones. Bottom: the mass-radius and mass-luminosity diagrams with linear fits through catalog data for primary and secondary components (blue lines with shading) and the relations for main sequence single stars (red lines). See text for references.}
\label{figRelAbs}
\includegraphics[width=\textwidth]{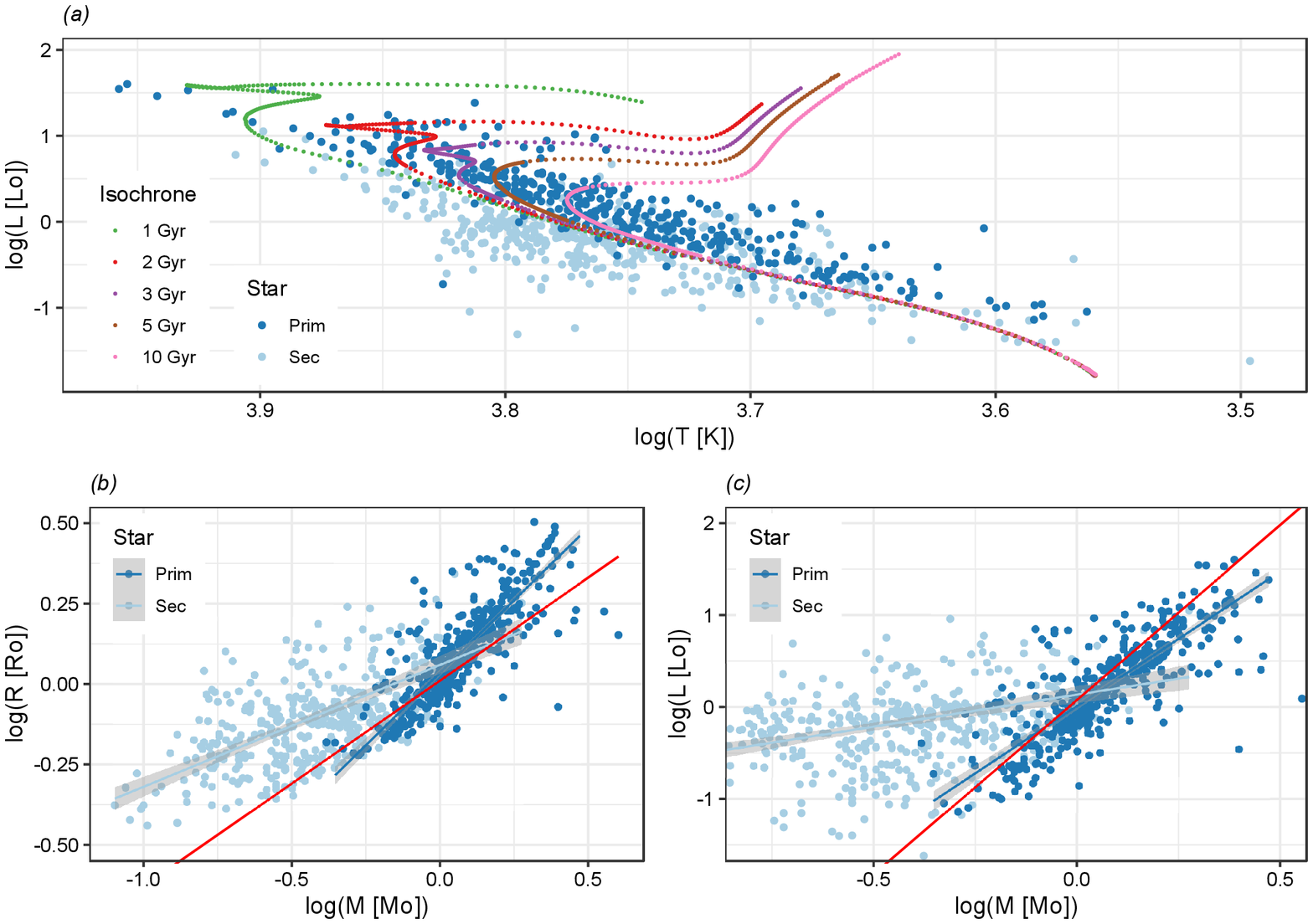}
\end{figure}

Fig. \ref{figRelAbs} shows the primary and secondary components of the cataloged systems on the HR diagram (panel \textit{a}), mass-radius (\textit{b}) and mass-luminosity diagram (\textit{c}). On the HR diagram are also shown the MIST isochrones for single stars \citep{MIST1, MIST2} with metallicity of $[Fe/H]=-0.25$ (informed by the statistics presented by \citealp{lamost}) and the rotation velocity $v/v_{crit}=0.4$ (because stars in close binaries rotate at higher rates than their single counterparts due to the spin-orbit synchronization). The red lines in the mass-radius and mass-luminosity diagrams show the relations for main sequence single stars \citep{allen}. The relations derived from our data (using again only the reliable sample), are given by the following equations:

\begin{eqnarray}
\label{eqRelMRML}
    \log R_1 = (0.90 \pm 0.03) \log M_1 + (0.04 \pm 0.01) \\ 
    \log R_2 = (0.38 \pm 0.03) \log M_2 + (0.06 \pm 0.01) \\
    \log L_1 = (2.92 \pm 0.11) \log M_1 + (0.01 \pm 0.02) \\ 
    \log L_2 = (0.69 \pm 0.09) \log M_2 + (0.13 \mp 0.05)
\end{eqnarray}

From Fig. \ref{figRelAbs} we can see that the primary components of the cataloged stars occupy the region of unevolved low-mass single stars, while the secondaries are systematically oversized and overluminous for their masses. This behavior is an expected consequence of the energy exchange through the common envelope.

\begin{table}
\begin{center}
\caption{Subsamples used for fitting parameter relations}
\label{tabSamples}
\begin{tabular}{llr}
\hline
Eq. & Conditions & Sample \\
\hline
4      & $q < 0.5$                                                         & \qTok \\
5,  6  & $M_1 < 3\ M_{\odot} \land (QT = SP \lor ET = 1)$                  & \qMok \\ 
7,  8  & $R_1 < 3\ R_{\odot} \land (QT = SP \lor ET = 1)$                  & \qRok \\
9, 10  & $L_1 < 5\ L_{\odot} \land (QT = SP \lor ET = 1)$                  & \qLok \\
11     & $P < 0.5\ d$                                                      & \PTok \\
12, 13 & $P < 0.5\ d \land M_1 < 2\ M_{\odot} \land (QT = SP \lor ET = 1)$ & \PMok \\
14, 15 & $P < 0.5\ d \land R_1 < 2\ R_{\odot} \land (QT = SP \lor ET = 1)$ & \PRok \\
16, 17 & $P < 0.5\ d \land L_1 < 5\ L_{\odot} \land (QT = SP \lor ET = 1)$ & \PLok \\
18, 19 & $QT = SP \lor ET = 1$                                             & \MRok \\
20, 21 & $QT = SP \lor ET = 1$                                             & \MLok \\
\hline
\end{tabular}
\tablecomments{The entire sample contains \objCount stars. Those in the ``reliable'' subsample are defined as either having a spectroscopic mass ratio ($QT=SP$) or a total eclipse ($ET=1$). The subsamples are further limited by how many stars have the required measurements, and finally by rejecting outliers with measurements outside the indicated range.}
\end{center}
\end{table}

%
%

\section{Summary and Conclusions}
\label{secSumCon}

Compiling a large sample of \objCount W UMa individually studied stars from the literature allowed us to conduct various statistics and formulate the following conclusions:

\begin{itemize}
\item There are hardly any objects with reliable parameters (as defined in Section \ref{secRel}) in our sample that have mass ratios above 0.5. Larger values derived by means of photometric modeling alone, without the support of a radial-velocity fit, should be treated as tentative at best.
\item Stars with periods longer than 0.5 days and temperatures higher than 7000 K form a separate group on the period-temperature plane. These stars likely have partially or completely radiative envelopes and should not be labeled as W UMa binaries.
\item The differences between the parameter distributions of A- and W-type stars are much less dramatic when the original definition is used for type assignment instead of secondary indicators.
\item Dividing the sample according to A/W types and absence/presence of spots results in remarkably similar parameter distributions. However, the types do not seem to be correlated with spots.
\item We find no correlation between the type of the period variability (monotonic increase or decrease, or cyclic changes) and A/W types, spots, or the mass ratio.
\item We also find no correlation between the fillout and the temperature difference between the components, contrary to the expectation that large temperature differences signify marginal contact.
\end{itemize}

The recent studies of \citet{sun2020} and \citet{li2020} essentially confirm our findings related to the ranges of mass ratios and periods of W UMa stars. \citet{sun2020} performed automated modeling of nearly 3000 totally eclipsing contact binaries from CRTS \citep{crts}, while \citet{li2020} did a similar study with 380 contact systems found in the Kepler Catalog of Eclipsing Binaries \citep{kepler}. In both these works a ``long period cutoff'' at about 0.5 days and a ``large mass ratio cutoff'' at $q \approx$ 0.5 are clearly discernible. The other statistics reported in these studies are also in agreement with our results.

Apart from the findings enumerated above, we hope that our catalog of W UMa binaries itself might be of use to the astronomical community for comparisons with different parameter estimation methods, statistical analyses we have not thought to conduct, or as a training set for machine learning.

%
%

\section{Data availability}
\label{secDataAv}

The entire catalog is available in the machine readable format (see Table \ref{tabColumns}). It is also available in the form of an interactive online database at \textit{https://wumacat.aob.rs}.

\acknowledgments

We thank M. Y{\i}ld{\i}z for helpful discussions about the age estimation method. The research presented in this paper was funded by the Ministry of Education, Science and Technological Development of Republic of Serbia through Grant No. 451-03-9/2021-14/200002. We gratefully acknowledge the use of the Simbad database ({\it http://simbad.u-strasbg.fr/simbad/}), operated at the CDS, Strasbourg, France, and NASA's Astrophysics Data System Bibliographic Services ({\it http://adsabs.harvard.edu/}).

%
%

\appendix


\section{Age Calculation}
\label{apxAgeCalc}

The procedure of age estimation of W UMa stars based on their masses and luminosities was developed by \citet{YD2013} (hereafter P13) and \citet{YI2014} (hereafter P14). A few key details were omitted in those papers that we later learned about in private communication. We outline the full procedure here (without repeating the explanations) to make our own results reproducible.

First, we calculate the quantity $M_L$, defined in Section 3 (and indirectly in Eq. 2) of P13 as:

\begin{equation}
M_L = \left(\frac{L_2}{1.49}\right)^\frac{1}{4.216}
\end{equation}

This is the mass of a main sequence star with the luminosity of the secondary in a W UMa system. Next we calculate the difference between $M_L$ and the actual mass of the secondary, $M_2$, as $\delta M = M_L-M_2$. We can then find the initial mass of the secondary, $M_{2i}$, from Eq. 7 in P13:

\begin{equation}
M_{2i} = M_2 + 2.5 (\delta M - 0.07)^{0.64}
\end{equation}

\noindent and the reciprocal value of the initial mass ratio, $1/q_i$, from Eq. 6 in P13:

\begin{equation}
\label{eqRecQi}
\frac{1}{q_i}=\frac{M_1 - (M_{2i}-M_2)(1-\gamma)}{M_{2i}}
\end{equation}

\noindent The quantity $\gamma$ is related to mass loss and it is defined in Eq. 5 of P13. A value of $\gamma=0.664$ is adopted in Section 4.2 of P13. 

From the initial mass ratio, we can now also calculate the initial mass of the primary, as $M_{1i}=(1/q_i)M_{2i}$.

The authors of P13 and P14 found that the values for the initial mass ratio calculated as above are correlated with the mass difference, $\delta M$, for W-type stars (whereas no correlation was found for A-type stars). They fit the correlation with the following exponential trend:

\begin{equation}
\frac{1}{q_i}=2.80-3.05\delta M^{0.25}
\end{equation}

The correlation is then accounted for by subtracting this trend from the reciprocal value of the initial mass ratio of affected objects. In this work, the correction is applied to all W-type stars where $\delta M<0.35$ using the expression:

\begin{equation}
\left(\frac{1}{q_i}\right)_{corr}=\frac{1}{q_i} - (2.80-3.05\delta M^{0.25}) + 0.5
\end{equation}

The last term is an additive constant required to align the mean of the corrected initial mass ratios with those unaffected by the trend. In the sample used in P13 and P14, the mean of the reciprocal value of the initial mass ratio for A-type stars is $0.492$, but the value of $0.5$ gave better results during our efforts to replicate the ages published in P13 and P14 using the sample provided there.

P13 and P14 do not state explicitly how the corrected reciprocal value of the initial mass ratios is used. In this work, it is used to recalculate the initial mass of the secondary by substituting it into Eq. \ref{eqRecQi}, so that, for the affected objects:

\begin{eqnarray}
(M_{2i})_{corr}=\frac{M_1+(1-\gamma)M_2}{(1/q_i)_{corr}+(1-\gamma)} \\
(M_{1i})_{corr}=(1/q_i)_{corr}(M_{2i})_{corr}
\end{eqnarray}

The lifetime of the star on the main sequence is calculated from its mass, using the following expression (Eq. 3 in P14):

\begin{equation}
t_{MS}=\frac{10}{M^{4.05}}\Big(0.0056(M + 3.993)^{3.16} + 0.042\Big)
\end{equation}

\noindent Here $M$ can be $M_{1i}$ for the primary, or $M_{2i}$ for the secondary component. Masses are expressed in solar units. Finally, the age of the system is given by:

\begin{equation}
t = 
\begin{cases}
t_{MS}(M_{2i})+t_{MS}(\overline{M_2}), & \text{for A-type stars (Eq. 6 in P14)} \\
t_{MS}(M_{2i}), & \text{for W-type stars (Eq. 8 in P14)}
\end{cases}
\end{equation}

\noindent where $\overline{M_2}=0.5(M_L+M_{2i})$ is the mean mass of the secondary during binary evolution (Eq. 4 from P14). In cases where the age calculated this way turns out to be greater than the main sequence lifetime of the primary component, $t_{MS}(M_{1i})$, the main sequence lifetime of the primary component is instead adopted as the age of the system.

\end{document}